\documentclass[journal]{IEEEtran}
\newcounter{mytempeqncnt}

\usepackage{booktabs}
\usepackage[noadjust]{cite}

\newcommand{\qed}{\nobreak \ifvmode \relax \else
      \ifdim\lastskip<1.5em \hskip-\lastskip
      \hskip1.5em plus0em minus0.5em \fi \nobreak
      \vrule height0.75em width0.5em depth0.25em\fi}

\ifCLASSINFOpdf
  \usepackage[pdftex]{graphicx}
  \graphicspath{{./}{Figs/}}
  \DeclareGraphicsExtensions{.pdf,.jpeg,.png}
\else
\fi
\usepackage{amssymb}
\usepackage{amsmath}
\usepackage{mathtools}
\usepackage{bigdelim}
\usepackage{tikz}
\usepackage{bbm}
\usepackage{mathtools}
\usepackage{centernot}
\usepackage{enumerate}
\usepackage{subcaption}
\usepackage{undertilde}
\usepackage{verbatim}
\usetikzlibrary{arrows,calc}
\usepackage{relsize}

\DeclarePairedDelimiterX{\infdivx}[2]{(}{)}{%
  #1\;\delimsize\|\;#2%
}

\newcommand{\defeq}{\vcentcolon=}

\begin{document}
\title{Bayesian Nonparametric Causal Inference: Information Rates and Learning Algorithms}

\author{Ahmed~M.~Alaa,~\IEEEmembership{Member,~IEEE}, and Mihaela van der Schaar,~\IEEEmembership{Fellow,~IEEE}
\thanks{A. Alaa is with the Department of Electrical Engineering, University of California Los Angeles (UCLA), Los Angeles, CA, 90095, USA (e-mail: ahmedmalaa@ucla.edu).} 
\thanks{M. van der Schaar is with the Department of Engineering Science, University of Oxford, Parks Road, Oxford, OX1 3PJ, UK (e-mail: mihaela.vanderschaar@eng.ox.ac.uk).} 
\thanks{Section \ref{mtgp} of this work was presented in part at the thirty-first annual conference on Neural Information Processing Systems (NIPS), 2017. This work was supported by the ONR Data science program.}
}
\markboth{XXXXXXXX, ~Vol.~XX, No.~X, XXXX~2018}%
{Alaa \MakeLowercase{\textit{et al.}}: Bayesian Nonparametric Causal Inference}
\maketitle
\begin{abstract} 
We investigate the problem of estimating the causal effect of a treatment on {\it individual} subjects from observational data; this is a central problem in various application domains, including healthcare, social sciences, and online advertising. Within the {\it Neyman-Rubin potential outcomes} model, we use the Kullback-Leibler (KL) divergence between the estimated and true distributions as a measure of accuracy of the estimate, and we define the information rate of the Bayesian causal inference procedure as the (asymptotic equivalence class of the) expected value of the KL divergence between the estimated and true distributions as a function of the number of samples. Using Fano's method, we establish a fundamental limit on the information rate that can be achieved by {\it any} Bayesian estimator, and show that this fundamental limit is independent of the {\it selection bias} in the observational data. We characterize the Bayesian priors on the potential (factual and counterfactual) outcomes that achieve the optimal information rate.  As a consequence, we show that a particular class of priors that have been widely used in the causal inference literature cannot achieve the optimal information rate. On the other hand, a broader class of priors can achieve the optimal information rate.  We go on to propose a prior adaptation procedure (which we call the {\it information-based empirical Bayes} procedure) that optimizes the Bayesian prior by maximizing an information-theoretic criterion on the recovered causal effects rather than maximizing the marginal likelihood of the observed (factual) data. Building on our analysis, we construct an information-optimal Bayesian causal inference algorithm. This algorithm embeds the potential outcomes in a {\it vector-valued reproducing kernel Hilbert space} (vvRKHS), and uses a multi-task Gaussian process prior over that space to infer the individualized causal effects. We show that for such a prior, the proposed information-based empirical Bayes method adapts the smoothness of the {\it multi-task Gaussian process} to the true smoothness of the causal effect function by balancing a tradeoff between the {\it factual bias} and the {\it counterfactual variance}. We conduct experiments on a well-known real-world dataset and show that our model significantly outperforms the state-of-the-art causal inference models.     
\end{abstract}
\IEEEpeerreviewmaketitle{ }
\section{Introduction}
\IEEEPARstart{T}{he} problem of estimating the {\it individualized} causal effect of a particular intervention from {\it observational data} is central in many application domains and research fields, including public health and healthcare \cite{foster2011subgroup}, computational advertising \cite{bottou2013counterfactual}, and social sciences \cite{xie2012estimating}. With the increasing availability of data in all these domains, machine learning algorithms can be used to obtain estimates of the effect of an intervention, an action, or a treatment on individuals given their features and traits. For instance, using observational electronic health record data\footnote{https://www.healthit.gov/sites/default/files/briefs/}, machine learning-based recommender system can learn the {\it individual-level} causal effects of treatments currently deployed in clinical practice and help clinicians refine their current treatment policies \cite{sauerbrei2014strengthening}. There is a growing interest in using machine learning methods to infer the individualized causal effects of medical treatments; this interest manifests in recent initiatives such as STRATOS \cite{sauerbrei2014strengthening}, which focuses on guiding observational medical research, in addition to various recent works on causal effect inference by the machine learning community \cite{alaa2017bayesian,tran2016model,wager2017estimation,hill2011bayesian,li2017matching,kallus2017recursive}.

The problem of estimating individual-level causal effects is usually formulated within the classical {\it potential outcomes} framework, developed by Neyman and Rubin \cite{rosenbaum1984reducing,rubin1974estimating}. In this framework, every subject (individual) in the observational dataset possesses two ``potential outcomes": the subject's outcome under the application of the treatment, and the subject's outcome when no treatment is applied. The treatment effect is the difference between the two potential outcomes, but since we only observe the ``factual" outcome for a specific treatment assignment, and never observe the corresponding ``counterfactual" outcome, we never observe any samples of the true treatment effect in an observational dataset. {\it This is what makes the problem of causal inference fundamentally different from standard supervised learning (regression)}. Moreover, the policy by which treatments are assigned to subjects induces a {\it selection bias} in the observational data, creating a discrepancy in the feature distributions for the {\it treated} and {\it control} patient groups, which makes the problem even harder. Many of the classical works on causal inference have focused on the simpler problem of estimating {\it average} treatment effects, where unbiased estimators based on propensity score weighting were developed to alleviate the impact of selection bias on the causal estimands (see \cite{abadie2016matching} and the references therein). 

While more recent works have developed machine learning algorithms for estimating individualized treatment effects from observational data in the past few years \cite{alaa2017bayesian,Onur1,Alaa1,hahn2017bayesian,bottou2013counterfactual,chernozhukov2016double,hill2011bayesian,johansson2016learning,powers2017some}, the inference machinery built in most of these works seem to be rather ad-hoc. The causal inference problem entails a richer set of modeling choices and decisions compared to that of the standard supervised learning (regression) problem, which includes deciding what model to use, how to model the treatment assignment variables in the observational data, and how to handle selection bias, etc. In order to properly address all these modeling choices, one needs to understand the fundamental limits of performance in causal effect estimation problems, and how different modeling choices impact the achievable performance.    

In this paper, we establish the fundamental limits on the amount of information that a learning algorithm can gather about the causal effect of an intervention given an observational data sample. We also provide guidelines for building proper causal inference models that ``do not leave any information on the table" because of poor modeling choices. A summary of our results is provided in the following Section. 

\section{Summary of the Results}
We address the individualized causal effect estimation problem on the basis of the Neyman-Rubin potential outcomes model \cite{rosenbaum1984reducing,rubin1974estimating}. We focus on Bayesian nonparametric learning algorithms, as they are immune to model mis-specification, and can learn highly heterogeneous response functions that one would expect to encounter in datasets with medical or social outcomes \cite{xie2013population,xie2012estimating}. In Section \ref{secIII}, we introduce the notion of {\it information rate} as a measure for the quality of Bayesian nonparametric learning of the individualized causal effects. The information rate is defined in terms of a measure of the Kullback-Leibler divergence between the true and posterior distributions for the causal effect. In Theorem 1, we establish the equivalence between Bayesian information rates and frequentist estimation rate. In the rest of the paper, we characterize: (1) the optimal information rates that can be achieved by any Bayesian nonparametric learning algorithm, and (2) the nature of the priors that would give rise to ``informationally optimal" Bayesian nonparametric causal inference procedure. 

In Section \ref{secIV}, we establish the fundamental limit on the information rate that can be achieved by any Bayesian causal inference procedure using an information-theoretic lower bound based on Fano's method. The optimal information rate is a property of the function classes to which the potential outcomes belong, and is independent of the inference algorithm. We show that the optimal information rate for causal inference is governed by the ``rougher" of the two potential outcomes functions. We also show that the optimal information rates for causal inference are {\it insensitive} to selection bias (Theorem 2).

In Section \ref{secadapt}, we characterize the Bayesian priors that achieve the optimal rate. We show that the most common modeling choice adopted in the literature, which is to augment the treatment assignment variable to the feature space, leads to priors that are suboptimal in terms of the achievable rate (Theorem 3). We show that informationally optimal priors are ones that place a probability distribution over a vector-valued function space, where the function space has its smoothness matching the rougher of the two potential outcomes functions. Since the true smoothness parameter of the potential outcomes functions is generally unknown a priori, we propose a prior adaptation procedure, called the {\it information-based empirical Bayes} procedure, which optimizes the Bayesian prior by maximizing an information-theoretic criterion on the recovered causal effects rather than maximizing the marginal likelihood of the observed (factual) data.       

We conclude the paper by building an information-optimal Bayesian causal inference algorithm that is based on our analysis. The inference procedure embeds the potential outcomes in a {\it vector-valued reproducing kernel Hilbert space} (vvRKHS), and uses a {\it multi-task Gaussian process} prior (with a Mat\'ern kernel) over that space to infer the individualized causal effects. We show that for such a prior, the proposed information-based empirical Bayes method exhibits an insightful {\it factual bias} and {\it counterfactual variance} decomposition. Experiments conducted on a standard dataset that is used for benchmarking causal inference models show that our model significantly outperforms the state-of-the-art.        

\section{Related Work}
\label{secII}
We conduct our analysis within the {\it potential outcomes} framework developed by Neyman and Rubin \cite{rosenbaum1984reducing,rubin1974estimating}. The earliest works on estimating causal effects have focused on the problem of obtaining unbiased estimates for the {\it average} treatment effects using observational samples. The most common well-known estimator for the average causal effect of a treatment is the propensity score weighting estimator, which simply removes the bias introduced by selection bias by giving weights to different samples that are inversely proportional to their propensity scores \cite{abadie2016matching}. More recently, the machine learning community has also developed estimators for the average treatment effects that borrows ideas from representation learning, i.e. see for instance the work in \cite{li2017matching}. In this paper, we focus on the {\it individual}, rather than the {\it average} causal effect estimation problem. 

To the best of our knowledge, non of the previous works have attempted to characterize the limits of learning causal effects in either the frequentist or Bayesian setups. Instead, most previous works on causal effect inference have focused on model development, and various algorithms have been recently developed for estimating individualized treatment effects from observational data, mostly based on either tree-based methods \cite{hahn2017bayesian,hill2011bayesian,wager2017estimation}, or deep learning methods \cite{Onur1,Alaa1}. Most of the models that were previously developed for estimating causal effects relied on regression models that treat the treatment assignment variables (i.e. whether or not the intervention was applied to the subject) as an extended dimension in the feature space. Examples of such models include Bayesian additive regression trees (BART) \cite{hill2011bayesian}, causal forests \cite{wager2017estimation}, balanced counterfactual regression \cite{johansson2016learning}, causal multivariate additive regression splines (MARS) \cite{powers2017some}, propensity-dropout networks \cite{Alaa1}, or random forests \cite{lu2017estimating}. In all these methods, augmenting the treatment assignment variable to the feature space introduces a mismatch between the training and testing distribution (i.e. covariate shift induced by the selection bias \cite{johansson2016learning}). The different methods followed different approaches for handling the selection bias: causal forests use estimates of the propensity score for deriving a tree splitting rule that attempts to balance the treated and control populations, propensity-dropout networks use larger dropout regularization for training points with very high or very low propensity scores, whereas balanced counterfactual regression uses deep neural networks to learn a balanced representation (i.e. a feature transformation) that tries to alleviate the effect of the selection bias. Bayesian methods, like BART, do not address selection bias since the Bayesian posterior naturally incorporates uncertainty in regions of poor overlap in the feature space. As we show later in Sections \ref{secadapt} and \ref{expsec}, our analysis and experimental results indicated that, by augmenting the treatment assignment variable to the feature space, all these methods achieve a suboptimal information rate.        

Our analysis is related to a long strand of literature that studied frequentist (minimax) estimation rates, or posterior contraction rates in standard regression problems \cite{rockova2017posterior,linero2017bayesian,sniekers2015adaptive,stone1982optimal,stone1980optimal}. In Theorem 2, we show that the optimal information rate for causal inference has the same form as the optimal minimax estimation rate obtained by Stone in \cite{stone1982optimal} for standard nonparametric regression problems, when the true regression function is set to be the rougher of the two potential outcomes functions. Our analysis for the achievable information rates for Gaussian process priors uses the results by van Zanten and van der Vaart in \cite{vaart2011information}.    

\section{Bayesian Nonparametric Causal Inference\\ from Observational Data}
In this section, we provide a general description for the Neyman-Rubin causal model considered in this paper (Subsection \ref{secIIIA}), and present the Bayesian nonparametric inference framework under study (Subsection \ref{secIIIB}). 
\label{secIII}
\subsection{The Neyman-Rubin Causal Model}
\label{secIIIA}
Consider a population of subjects with each subject $i$ possessing a $d$-dimensional {\it feature} $X_i \in \mathcal{X}$. An intervention is applied to some subjects in the population: subject $i$'s response to the intervention is a random variable denoted by $Y^{(1)}_i$, whereas the subject's natural response when no intervention is applied is denoted by $Y^{(0)}_i$. The two random variables, $Y^{(1)}_i, Y^{(0)}_i \in \mathbb{R}$, are known as the {\it potential outcomes}. The causal effect of the intervention (treatment) on subject $i$ is characterized through the difference between the two (random) potential outcomes $(Y^{(1)}_i-Y^{(0)}_i)\,|\,X_i = x$, and is generally assumed to be dependent on the subject's features $X_i = x$. Hence, we define the {\it individualized treatment effect} (ITE) for a subject $i$ with a feature $X_i = x$ as    
\begin{equation}
T(x) = \mathbb{E}\left[\left. Y^{(1)}_i-Y^{(0)}_i\,\right|\,X_i = x\right].
\label{eq1}
\end{equation}
Our goal is to estimate the function $T(x)$ from an {\it observational} dataset $\mathcal{D}_n$, which comprises $n$ independent samples of the random tuple $\{X_i, \omega_i, Y^{(\omega_i)}_i\},$ where $\omega_i \in \{0,1\}$ is an intervention assignment indicator that indicates whether or not subject $i$ has received the intervention (treatment) under consideration. The outcomes $Y^{(\omega_i)}_i$ and $Y^{(1-\omega_i)}_i$ are known in the literature as the {\it factual} and the {\it counterfactual} outcomes, respectively \cite{johansson2016learning}, \cite{shalit2016estimating}. Intervention assignments generally depend on the subjects' features, i.e. $\omega_i \not\!\perp\!\!\!\perp X_i$. This dependence is quantified via the conditional distribution $\mathbb{P}(\omega_i=1|X_i=x)$, also known as the {\it propensity score} of subject $i$ \cite{abadie2016matching}, \cite{rosenbaum1984reducing}. In the rest of this paper, we denote the propensity score of a feature point $x$ as $\gamma(x)$.    

The observational dataset $\mathcal{D}_n = \{X_i, \omega_i, Y^{(\omega_i)}_i\}^n_{i=1}$ is drawn from a joint density $d\mathbb{P}(X_i,\omega_i,Y^{(0)}_i,Y^{(1)}_i),$ with a probability space $(\Omega,\mathcal{F},\mathbb{P})$ that supports the following standard conditions \cite{rosenbaum1984reducing,rubin1974estimating}:
\begin{itemize}
\item {\bf Condition 1 (unconfoundedness):} Treatment assignment decisions are independent of the outcomes given the subject's features, i.e. $(Y^{(0)}_i,Y^{(1)}_i) \!\perp\!\!\!\perp \omega_i\,|\,X_i$.
\item {\bf Condition 2 (overlap):} Every subject has a non-zero chance of receiving the treatment, and treatment assignment decisions are non-deterministic, i.e. $0<\gamma(x)<1$.
\end{itemize}  
\subsection{Bayesian Nonparametric Causal Inference}
\label{secIIIB}

Throughout this paper, we consider the following {\it signal-in-white-noise} random design regression model for the potential outcomes:
\begin{equation}
Y^{(\omega)}_i = f_{\omega}(X_i) + \epsilon_{i,\omega},\, \omega \in \{0,1\},  
\label{eq2}
\end{equation}
where $\epsilon_{i,\omega} \sim \mathcal{N}(0,\sigma^2_{\omega})$ is a Gaussian noise variable. It follows from (\ref{eq2}) that $\mathbb{E}[Y^{(\omega)}_i\,|\, X_i = x] = f_{\omega}(x),$ and hence the ITE is given by $T(x) = f_{1}(x)-f_{0}(x)$. The functions $f_{1}(x)$ and $f_{0}(x)$ correspond to the {\it response surfaces} over the subjects' feature space with and without the intervention; the difference between these two surfaces correspond to the individualized effect of the intervention. We assume that $\mathcal{X}$ is a compact metric space (e.g. bounded, closed sets in $\mathbb{R}^d$), and that the true response surfaces $f_{\omega}: \mathcal{X} \rightarrow \mathbb{R}$, $\omega \in \{0,1\}$ are totally bounded functions in a space of ``smooth" or ``regular" functions $\mathcal{\mathcal{F}}^{\alpha_{\omega}}$, where $\alpha_{\omega}$ is a smoothness (or regularity) parameter. This roughly means that $f_{\omega}$ is $\alpha_{\omega}$-differentiable; precise definitions for $\alpha_{\omega}$-regular function classes will be provided in subsequent Sections.

A Bayesian procedure for estimating the ITE function entails specifying a prior distribution $\Pi$ over the response surfaces $f_1(x)$ and $f_0(x)$, which in turn induces a prior over $T(x)$. The nonparametric nature of inference follows from the fact that $\Pi$ is a prior over functions, and hence the estimation problem involves an infinite-dimensional parameter space. For a given prior $\Pi$, the Bayesian inference procedure views the observational dataset $\mathcal{D}_n$ as being sampled according to the following generative model: 
\begin{align} 
f_0, f_1 &\sim \Pi,\,\, X_i \sim d\mathbb{P}(X_i=x) \nonumber \\
\omega_i\,|\,X_i=x &\sim \mbox{Bernoulli}(\gamma(x)) \nonumber \\
Y^{(\omega_i)}_i\,|\,f_0, f_1,\omega_i &\sim \mathcal{N}(f_{\omega_i}(x), \sigma^2_{\omega_i}),\, i = 1,.\,.\,.,n.
\label{eq3}
\end{align}
Since we are interested in estimating an underlying true ITE function $T(x)$, we will analyze the Bayesian causal inference procedure within the {\it frequentist} setup, which assumes that the subjects' outcomes $\{Y^{(\omega_i)}_i\}^n_{i=1}$ are generated according to the model in (\ref{eq3}) for a given true (and fixed) regression functions $f_0(x)$ and $f_1(x)$. That is, in the next Subsection, we will assess the quality of a Bayesian inference procedure by quantifying the amount of information the posterior distribution $d\Pi_n(T\,|\,\mathcal{D}_n) = d\Pi_n(f_1-f_0\,|\,\mathcal{D}_n)$ has about the true ITE function $T$. This type of analysis is sometimes referred to as the ``Frequentist-Bayes" analysis \cite{van1998asymptotic}.

\subsection{Information Rates}
\label{secIIIC} 
How much information about the true causal effect function $T(.)$ is conveyed in the posterior $d\Pi_n(T\,|\,\mathcal{D}_n)$? A natural measure of the ``informational quality" of a posterior $d\Pi_n(T\,|\,\mathcal{D}_n)$ is the information-theoretic criterion due to Barron \cite{yang1999information}, which quantifies the quality of a posterior via the Kullback-Leibler (KL) divergence between the posterior and true distributions. In that sense, the quality (or informativeness) of the posterior $d\Pi_n(T\,|\,\mathcal{D}_n)$ at a feature point $x$ is given by the KL divergence between the posterior distribution at $x$, $d\Pi_n(T(x)\,|\,\mathcal{D}_n)$, and the true distribution of $(Y^{(1)}-Y^{(0)})\,|\,X = x$. The overall quality of a posterior is thus quantified by marginalizing the pointwise KL divergence over the feature space $\mathcal{X}$. For a prior $\Pi$, true responses $f_0$ and $f_1$, propensity function $\gamma$, and observational datasets of size $n$, the {\it expected KL risk} is:     
\begin{align} 
\mathbb{D}_n(\Pi; f_0, f_1,\gamma) = \mathbb{E}_{x}\left[\,\mathbb{E}_{\mathcal{D}_n}\left[D_{\mathrm {\small KL}}\left(P(x)\,\left\|\,Q_{\mathcal{D}_n}(x)\right)\right.\right]\,\right], 
\label{eq4}
\end{align}
where $D_{\mathrm {KL}}(.\|.)$ is the KL divergence\footnote{The KL divergence between probability measures $P$ and $Q$ is given by $D_{\mathrm {KL}}(P\|Q) = \int \log(dP/dQ) dP$ \cite{cover2012elements}. The existence of the Radon-Nikodym derivative $\frac{dP}{dQ_n}$ in (\ref{eq4}) is guaranteed since $P$ and $Q$ are mutually absolutely continuous.},~$P(x)$~is~the~true~distribution~of $T(x)$, i.e. $P(x) = d\mathbb{P}(Y^{(1)}-Y^{(0)}\,|\,X = x)$, and $Q_{\mathcal{D}_n}(x)$ is the posterior distribution of $T(x)$, given by:
\begin{align} 
Q_{\mathcal{D}_n}(x) &= d\Pi_n(Y^{(1)}-Y^{(0)}\,|\,X = x, \mathcal{D}_n) \nonumber \\
&\overset{\small (\star)}{=} d\Pi_n(T(x)+\mathcal{N}(0,\sigma^2_0+\sigma^2_1)\,|\,\mathcal{D}_n), \nonumber \\
&\overset{\small (*)}{=} \int \mathcal{N}(T(x),\sigma^2_0+\sigma^2_1) \, d\Pi_n(T(x)\,|\,\mathcal{D}_n),
\label{eq5}
\end{align}
where steps $(\star)$ and $(*)$ in (\ref{eq5}) follow from the sampling model in (\ref{eq3}). The expected KL risk $\mathbb{D}_n$ in (\ref{eq4}) marginalizes the pointwise KL divergence $D_{\mathrm {\small KL}}(P(x)\,\|\,Q_{\mathcal{D}_n}(x))$ over the distribution of the observational dataset $\mathcal{D}_n$ (generated according to (\ref{eq3})), and the feature distribution $d\mathbb{P}(X=x)$. Variants of the expected KL risk in (\ref{eq4}) have been widely used in the analysis of nonparametric regression models, usually in the form of (cumulative) Ces\`aro averages of the pointwise KL divergence at certain points in the feature space \cite{seeger2008information,yang1999information,bernardo1998information,vaart2011information}. Assuming posterior consistency, the (asymptotic) dependence of $\mathbb{D}_n(\Pi; f_0, f_1, \gamma)$ on $n$ reflects the rate by which the posterior $d\Pi_n(T\,|\,\mathcal{D}_n)$ ``sandwiches" the true ITE function $T(x)$ everywhere in $\mathcal{X}$. An efficient causal inference procedure would exhibit a rapidly decaying $\mathbb{D}_n$: this motivates the definition of an ``{\it information rate}".\\
\\   
{\bf Definition 1. (Information Rate)} We say that the {\it information rate} of a Bayesian causal inference procedure is $\mathcal{I}_n$, for a sequence $\mathcal{I}_n \to 0$, if $\mathbb{D}_n(\Pi; f_0, f_1, \gamma)$ is $\Theta(\mathcal{I}_n)$.\,\, \IEEEQED\\
\\   
Note~that~$\mathcal{I}_n$~is~the~equivalence~class~of~all~sequences~that have the same asymptotic rate of convergence. In the rest of this paper, we use the notation $\mathcal{I}_n(\Pi; f_0, f_1,\gamma)$ to denote the information rate achieved by a prior $\Pi$ in a causal inference problem instance described by the tuple $(f_0, f_1,\gamma)$. The notion of an ``information rate" for a Bayesian causal effect inference procedure is closely connected to the frequentist estimation rate (with respect to the $L_2$ loss) with $T(.)$ as the estimand \cite{yang1999information,chen2017convex,yang2015minimax}. The following Theorem establishes such a connection.\\    
\\
{\bf Theorem~1.}~Let~$\mathbb{D}_n(\Pi; f_0, f_1,\gamma)$~be~the~KL~risk~of~a~given Bayesian causal inference procedure, then we have that
\begin{align} 
\mathbb{D}_n(\Pi; f_0, f_1,\gamma) \leq \bar{\sigma}\cdot \mathbb{E}_{\mathcal{D}_n}\left[\,\big\Vert\mathbb{E}_{\Pi}\left[\,T\,|\,\mathcal{D}_n\,\right]-T\big\Vert^2_2\,\right], \nonumber
\end{align}
for some $\bar{\sigma} > 0$, where $\|.\|_2^2$ is the $L_2(\mathbb{P})$-norm with respect to the feature distribution, i.e. $\|f\|_2^2 = \int f^2(x) d\mathbb{P}(X=x)$.\\
\\
{\bf Proof.} Recall from (\ref{eq4}) that $\mathbb{D}_n$ is given by
\begin{align} 
\mathbb{D}_n(\Pi; f_0, f_1,\gamma) = \mathbb{E}_{x}\left[\,\mathbb{E}_{\mathcal{D}_n}\left[D_{\mathrm {\small KL}}\left(P(x)\,\left\|\,Q_{\mathcal{D}_n}(x)\right)\right.\right]\,\right]. \nonumber
\end{align}
Based on (\ref{eq5}), $D_{\mathrm {\small KL}}\left(P(x)\,\left\|\,Q_{\mathcal{D}_n}(x)\right.\right)$ can be written as
\begin{align} 
D_{\mathrm {\small KL}}\left(P(x)\,\left\|\,\int \mathcal{N}(T(x),\sigma^2_0+\sigma^2_1) \, d\Pi_n(T(x)\,|\,\mathcal{D}_n)\right.\right), \nonumber
\end{align} 
which by the convexity of the KL divergence in its second argument, and using Jensen's inequality, is bounded above by
\begin{align} 
D_{\mathrm {\small KL}}\left(P(x)\,\left\|\, \mathcal{N}\left(\mathbb{E}_{\Pi}[\,T(x)\,|\,\mathcal{D}_n\,],\sigma^2_0+\sigma^2_1\right)\right.\right). \nonumber
\end{align}
From the regression model in (\ref{eq2}), we know that $P(x) = d\mathbb{P}(Y^{(1)}-Y^{(0)}\,|\,X = x) \sim \mathcal{N}(T(x),\sigma^2_0+\sigma^2_1)$, and hence KL divergence above can be written as  
\begin{align} 
D_{\mathrm {\small KL}}\left(\mathcal{N}(T(x),\sigma^2_0+\sigma^2_1)\,\left\|\, \mathcal{N}\left(\mathbb{E}_{\Pi}[\,T(x)\,|\,\mathcal{D}_n\,],\sigma^2_0+\sigma^2_1\right)\right.\right), \nonumber
\end{align}
which is given by $\frac{1}{2(\sigma^2_0+\sigma^2_1)}\left|\,\mathbb{E}_{\Pi}[\,T(x)\,|\,\mathcal{D}_n\,]-T(x)\,\right|^2$ since $D_{\mathrm {\small KL}}(\mathcal{N}(\mu_0,\sigma^2)\|\,\mathcal{N}(\mu_1,\sigma^2)) = \frac{1}{\sigma^2}|\mu_1-\mu_0|^2$ \cite{cover2012elements}. Hence, the expected KL risk is bounded above as follows
\begin{align} 
\mathbb{D}_n(\Pi; f_0, f_1,\gamma) &\leq \mathbb{E}_{\mathcal{D}_n}\left[\,\mathbb{E}_{x}\left[\frac{\left|\,\mathbb{E}_{\Pi}[\,T(x)\,|\,\mathcal{D}_n\,]-T(x)\,\right|^2}{2(\sigma^2_0+\sigma^2_1)}\right]\,\right], \nonumber \\
&= \frac{1}{2(\sigma^2_0+\sigma^2_1)}\,\mathbb{E}_{\mathcal{D}_n}\left[\,\Vert\,\mathbb{E}_{\Pi}[\,T\,|\,\mathcal{D}_n\,]-T\,\Vert_2^2\,\right], \nonumber
\end{align}
for all $n \in \mathbb{N}_+$.\,\, \IEEEQED\\ 
\\ 
Theorem 1 says that the information rate of causal inference lower bounds the rate of convergence of the $L_2(\mathbb{P})$ risk of the sequence of estimates $\hat{T}_n$ induced by the posterior mean $\int T d\Pi_n(T\,|\,\mathcal{D}_n)$. The $L_2(\mathbb{P})$ risk $\,\Vert\,\mathbb{E}_{\Pi}[\,T\,|\,\mathcal{D}_n\,]-T\,\Vert_2^2$ was dubbed the {\it precision in estimating heterogeneous effects} (PEHE) by Hill in \cite{hill2011bayesian}, and is the most commonly used metric for evaluating causal inference models \cite{athey2016recursive, hill2011bayesian, johansson2016learning, lu2017estimating, shalit2016estimating}. Theorem 1 tells us that the PEHE is $\Omega(\mathcal{I}_n)$, and hence the Bayesian information rate presents a limit on the achievable performance of frequentist estimation. In that sense, the asymptotic behavior of $\mathcal{I}_n(\Pi; f_0, f_1,\gamma)$ is revealing of both the informational quality of the Bayesian posterior, as well as the convergence rates of frequentist loss functions.

\section{Optimal Information Rates for\\ Bayesian Causal Inference}
\label{secIV} 
In this Section, we establish a fundamental limit on the information rate that can be achieved by {\it any} sequence of posteriors $d\Pi_n(T\,|\,\mathcal{D}_n)$ for a given causal inference problem. Let the {\it achievable information rate} for a given prior $\Pi$ and function classes $\mathcal{F}^{\alpha_0}$ and $\mathcal{F}^{\alpha_1}$, denoted by $I_n(\Pi; \mathcal{F}^{\alpha_0}, \mathcal{F}^{\alpha_1}, \gamma)$, be the rate obtained by taking the supremum of the information rate over functions in $\mathcal{F}^{\alpha_0}$ and $\mathcal{F}^{\alpha_1}$. This is a quantity that depends only on the prior but not on the specific realizations of $f_0$ and $f_1$. The {\it optimal information rate} is defined to be the maximum worst case achievable information rate for all functions in $\mathcal{F}^{\alpha_0}$ and $\mathcal{F}^{\alpha_1}$, and is denote by $I^*_n(\mathcal{F}^{\alpha_0}, \mathcal{F}^{\alpha_1}, \gamma)$. While the information rate $\mathcal{I}_n(\Pi; f_0, f_1, \gamma)$ characterizes a particular instance of a causal inference problem with $(f_0, f_1, \gamma)$ and a given Bayesian prior $\Pi$, the optimal information rate $I^*_n(\mathcal{F}^{\alpha_0}, \mathcal{F}^{\alpha_1}, \gamma)$ is an abstract (prior-independent) measure of the ``information capacity" or the ``hardness" of a class of causal inference problems (corresponding to response surfaces in $\mathcal{F}^{\alpha_0}$ and $\mathcal{F}^{\alpha_1}$). Intuitively, one expects that the limit on the achievable information rate will be higher for smooth (regular) response surfaces and for propensity functions that are close to 0.5 everywhere in $\mathcal{X}$. Theorem 2 provides a detailed characterization for the optimal information rates in general function spaces. Whether or not the Bayesian inference procedure achieves the optimal information rate will depend on the prior $\Pi$. In the next Section, we will investigate different design choices for the prior $\Pi$, and characterize the ``capacity-achieving" priors that achieve the optimal information rate.   

\begin{figure}[t]
\includegraphics[width=4in]{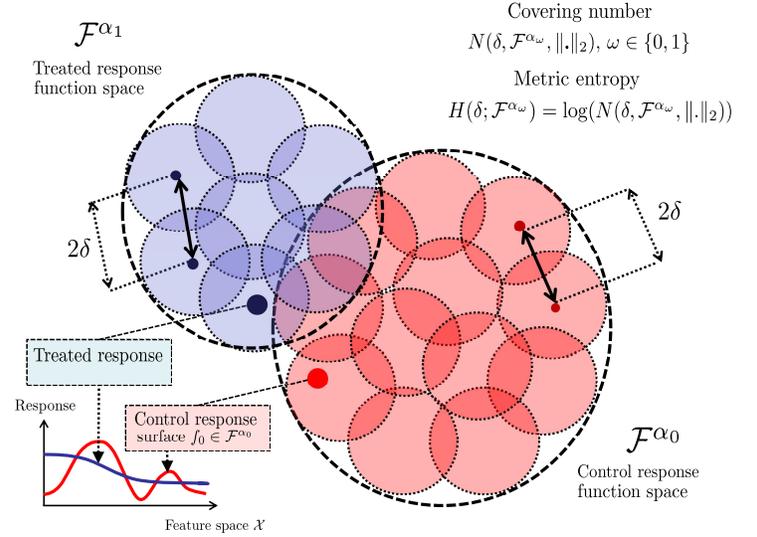}
\caption{\small Pictorial depiction of covering sets for $\mathcal{F}^{\alpha_0}$ and $\mathcal{F}^{\alpha_1}$.}
\label{fig2}
\end{figure}

In Theorem 2, we will use the notion of {\it metric entropy} $H(\delta; \mathcal{F}^{\alpha})$ to characterize the ``size" of general (nonparametric or parametric) function classes. The metric entropy $H(\delta; \mathcal{F}^{\alpha})$ of a function space $\mathcal{F}^{\alpha}$ is given by the logarithm of the {\it covering number} $N(\delta,\mathcal{F}^{\alpha},\rho)$ of that space with respect to a metric~$\rho$,~i.e.~$H(\delta; \mathcal{F}^{\alpha}) = \log(N(\delta,\mathcal{F}^{\alpha},\rho))$.~A~formal~definition for covering numbers is provided below.\\
\\   
{\bf Definition~2.~(Covering~number)}~A~$\delta$-cover~of~a~given~function space $\mathcal{F}^{\alpha}$ with respect to a metric $\rho$ is a set of functions $\{f^1,.\,.\,., f^N\}$ such that for any function $f \in \mathcal{F}^{\alpha}$, there exists some $v \in \{1,.\,.\,.,N\}$ such that $\rho(f, f^v) \leq \delta$. The $\delta$-covering number of $\mathcal{F}^{\alpha}$ is \cite{van1998asymptotic} 
\begin{align}
N(\delta,\mathcal{F}^{\alpha},\rho) \defeq \inf\{N \in \mathbb{N}: \mbox{$\exists$ a $\delta$-cover of $\mathcal{F}^{\alpha}$}\}. \,\,\,\,\, \IEEEQED \nonumber 
\end{align}
That is, the covering number of a function class $\mathcal{F}^{\alpha}$ is the number of balls (in a given metric $\rho$) of a fixed radius $\delta > 0$ required to cover it. Throughout this paper, the metric entropy will always be evaluated with respect to the $L_2(\mathbb{P})$ norm. In the light of the definition above, the metric entropy can be thought of as a measure of the complexity of a function class; smoother function classes would generally display a smaller metric entropy. All function classes considered in this paper have finite metric entropy. Figure \ref{fig2} shows a pictorial depiction for two exemplary function classes $\mathcal{F}^{\alpha_0}$ and $\mathcal{F}^{\alpha_1}$ for the treated and control responses, respectively. In this depiction, $\alpha_0$ is smaller than $\alpha_1$, hence the $\delta$-cover of $\mathcal{F}^{\alpha_0}$ contains more balls than the $\delta$-cover of $\mathcal{F}^{\alpha_1}$, and it follows that $\mathcal{F}^{\alpha_0}$ has a larger metric entropy. This manifests in the control response surface $f_0$ being less smooth than the treated response surface $f_1$. This is usually the case for real-world data on responses to medical treatments, where the untreated population typically display more heterogeneity than the treated population \cite{xie2013population}.
\begin{table*}
\centering
\caption{\small \textsc{Optimal information rates for Bayesian causal inference in standard function spaces.}}
\label{tab1}
\begin{tabular}{llll}
\toprule
{\small \textsc{Space}} & {\small \textsc{Metric Entropy Rate}} & {\small \textsc{Response surfaces}}& {\small \textsc{Optimal Information Rate}} \\
{\small $\mathcal{F}^{\alpha}$} & {\small $H(\delta; \mathcal{F}^{\alpha})$} & {\small $f_0,\,f_1$}& {\small $I^*_n(\mathcal{F}^{\alpha_0}, \mathcal{F}^{\alpha_1})$}\\ 
\midrule
Analytic $C^{\omega}(\mathcal{X})$ & $H(\delta; C^{\omega}) \asymp \log\left(\frac{1}{\delta}\right)$ & $f_0, f_1 \in C^{\omega}(\mathcal{X})$ & $\Theta(n^{-1})$ \\ 
Smooth $C^{\infty}(\mathcal{X})$ & $H(\delta; C^{\infty}) \asymp \log\left(\frac{1}{\delta}\right)$ & $f_0, f_1 \in C^{\infty}(\mathcal{X})$ & $\Theta(n^{-1})$ \\ 
$\alpha$-Smooth $C^{\alpha}(\mathcal{X})$ & $H(\delta; C^{\alpha}) \asymp \delta^{-\frac{d}{\alpha}}$ & $f_0 \in C^{\alpha_0}(\mathcal{X}),\, f_1 \in C^{\alpha_1}(\mathcal{X})$ & $\Theta\left(n^{-2(\alpha_0 \wedge \alpha_1)/(2(\alpha_0 \wedge \alpha_1)+d)}\right)$\\ 
H\"older $H^{\alpha}(\mathcal{X})$ & $H(\delta; H^{\alpha}) \asymp \delta^{-\frac{d}{\alpha}}$ & $f_0 \in H^{\alpha_0}(\mathcal{X}),\, f_1 \in H^{\alpha_1}(\mathcal{X})$ & $\Theta\left(n^{-2(\alpha_0 \wedge \alpha_1)/(2(\alpha_0 \wedge \alpha_1)+d)}\right)$\\
Sobolev $S^{\alpha}(\mathcal{X})$ & $H(\delta; S^{\alpha}) \asymp \delta^{-\frac{d}{\alpha}}$ & $f_0 \in S^{\alpha_0}(\mathcal{X}),\, f_1 \in S^{\alpha_1}(\mathcal{X})$ & $\Theta\left(n^{-2(\alpha_0 \wedge \alpha_1)/(2(\alpha_0 \wedge \alpha_1)+d)}\right)$\\
Besov $B^{\alpha}_{p,q}(\mathcal{X})$ & $H(\delta; B^{\alpha}) \asymp \delta^{-\frac{d}{\alpha}}$ & $f_0 \in B^{\alpha_0}_{p,q}(\mathcal{X}),\, f_1 \in B_{p,q}^{\alpha_1}(\mathcal{X})$ & $\Theta\left(n^{-2(\alpha_0 \wedge \alpha_1)/(2(\alpha_0 \wedge \alpha_1)+d)}\right)$\\
Lipschitz $L^{\alpha}(\mathcal{X})$ & $H(\delta; L^{\alpha}) \asymp \delta^{-\frac{d}{\alpha}}$  & $f_0 \in L^{\alpha_0}(\mathcal{X}),\, f_1 \in L^{\alpha_1}(\mathcal{X})$ & $\Theta\left(n^{-2(\alpha_0 \wedge \alpha_1)/(2(\alpha_0 \wedge \alpha_1)+d)}\right)$\\
Parametric models & $H(\delta; \Theta) \asymp K\cdot \log\left(\frac{1}{\delta}\right), |\Theta| = K$ & $f_\omega(\theta_\omega), \theta_\omega \in \Theta_\omega, |\Theta_\omega| = K_\omega, \omega \in \{0,1\}$ & $\Theta\left((K_0 \wedge K_1)^2\cdot n^{-1}\right)$ \\
\bottomrule
\end{tabular}
\end{table*}

We now present the main result of this Section. In the following Theorem, we provide a general characterization for the optimal information rates of Bayesian causal inference when the treated and control surfaces are known to belong to function classes $\mathcal{F}^{\alpha_{1}}$ and $\mathcal{F}^{\alpha_{0}}$.\\
\\
{\bf Theorem~2.~(Optimal~Information~Rates)}~Suppose~that~$\mathcal{X}$~is a compact subset of $\mathbb{R}^d$, and that Conditions 1-2 hold. Then the optimal information rate is $\Theta(\delta^2_0 \vee \delta^2_1),$ where $\delta_{\omega}$ is the solution for $H(\delta_{\omega};\, \mathcal{F}^{\alpha_{\omega}}) \asymp n \, \delta^2_{\omega},\, \omega \in \{0,1\}$. \\  
\\ 
{\bf Proof.} See Appendix A. \,\,\, \IEEEQED \\
\\
Theorem~2~characterizes~$I^*_n(\mathcal{F}^{\alpha_0}, \mathcal{F}^{\alpha_1},\gamma)$~in~terms~of~the~metric entropies $H(\delta;\, \mathcal{F}^{\alpha_{0}})$~and~$H(\delta;\, \mathcal{F}^{\alpha_{1}})$~for~general~function classes $\mathcal{F}^{\alpha_{0}}$ and $\mathcal{F}^{\alpha_{1}}$. We~used~the~local~{\it Fano method}~to derive~an~information-theoretic~lower~bound~on~the~information~rate that can be achieved by any estimator \cite{yang1999information}. The characterization in Theorem 2 implies that selection bias {\it has no effect} on the achievable information rate. (Thus, in the rest of the paper we drop the dependency on $\gamma$ when referring to $I^*_n$.) That is, as long as the overlap condition holds, selection bias does not hinder the information rate that can be achieved by a Bayesian causal inference procedure, and we can hope to find a good prior $\Pi$ that achieves the optimal rate of posterior contraction around the true ITE function $T(x)$ irrespective of the amount of bias in the data. Theorem 2 also says that the achievable information rate is bottle-necked by the more ``complex" of the two response surfaces $f_0$ and $f_1$. Hence, we cannot hope to learn the causal effect at a fast rate if either of the treated or the control response surfaces are rough, even when the other surface is smooth.  

The general characterization of the optimal information rates in Theorem 2 is cast into specific forms by specifying the regularity classes $\mathcal{F}^{\alpha_{0}}$ and $\mathcal{F}^{\alpha_{1}}$. Table \ref{tab1} demonstrates the optimal information rates for standard function classes, including analytic, smooth, H\"older \cite[Section 6.4]{yang1999information}, Sobolev \cite{birman1967piecewise}, Besov \cite[Section 6.3]{yang1999information}, and Lipschitz functions \cite{tikhomirov1993varepsilon,vosburg1966metric}. A rough description for the optimal information rates of all nonparametric function spaces ($\alpha$-smooth, H\"older, Sobolev, Besov, and Lipschitz) can be given as follows. If $f_0$ is $\alpha_0$-regular (e.g. $\alpha_0$-differentiable) and $f_1$ is $\alpha_1$-regular, then the optimal information rate for causal inference is 
\begin{align}
I^*_n(\mathcal{F}^{\alpha_0}, \mathcal{F}^{\alpha_1}) \asymp n^{\frac{-2(\alpha_0 \wedge \alpha_1)}{2(\alpha_0 \wedge \alpha_1)+d}},\label{OptInf1}
\end{align}  
where~$\asymp$~denotes~{\it asymptotic~equivalence},~i.e.~in~Bachmann-Landau~notation,~$g(x) \asymp f(x)$~if~$g(x) = \Theta(f(x))$.~That~is, the regularity parameter of the rougher response surface, i.e. $\alpha_0 \wedge \alpha_1$, dominates the rate by which any inference procedure can acquire information about the causal effect. This is because, if one of the two response surfaces is much more complex (rough) than the other (as it is the case in the depiction in Figure \ref{fig2}), then the ITE function $T(x)$ would naturally lie in a function space that is at least as complex as the one that contains the rough surface. Moreover, the best achievable information rate depends only on the smoothness of the response surfaces and the dimensionality of the feature space, and is independent of the selection bias. Due to the nonparametric nature of the estimation problem, the optimal information rate for causal inference gets exponentially slower as we add more dimensions to the feature space \cite{linero2017bayesian,yang2015minimax}. 

Note that in Theorem 2, we assumed that for the surfaces $f_0$ and $f_1$, all of the $d$ dimensions of $\mathcal{X}$ are {\it relevant} to the two response surfaces. Now assume that surfaces $f_0$ and $f_1$ have relevant feature dimensions in the sets $\mathcal{P}_0$ and $\mathcal{P}_1$, respectively, where $|\mathcal{P}_\omega|=p_\omega \leq d,\,\omega \in \{0,1\}$ \cite{raskutti2009lower}, then
\begin{align}
I^*_n(\mathcal{F}^{\alpha_0}_{\mathcal{P}_0}, \mathcal{F}^{\alpha_1}_{\mathcal{P}_1}) \asymp n^{\frac{-2\alpha_0}{2\alpha_0+p_0}} \vee n^{\frac{-2\alpha_1}{2\alpha_1+p_1}},
\label{OptInf2}
\end{align}  
where $\mathcal{F}^{\alpha_\omega}_{\mathcal{P}_\omega}$ denotes the space of functions in $\mathcal{F}^{\alpha_\omega}$ for which the relevant dimensions are in $\mathcal{P}_\omega$. In (\ref{OptInf2}), the rate is dominated by the more complex response surface, where ``complexity" here is manifesting as a combination of the number of relevant dimensions and the smoothness of the response over the those dimensions. One implication of (\ref{OptInf2}) is that the information rate can be bottle-necked by the smoother of the response surfaces $f_0$ and $f_1$, if such a response has more relevant dimensions in the feature space\footnote{A more general characterization of the information rate would consider the case when the responses have different smoothness levels on each of the $d$-dimensions. Unfortunately, obtaining such a characterization is technically daunting.}. More precisely, if $\alpha_0 < \alpha_1$, then the information rate can still be bottle-necked by the smoother surface $f_1$ as long as $p_1 > \frac{\alpha_1}{\alpha_0}\,p_0$.  

 
Since the optimal (Bayesian) information rate is a lower bound on the (frequentist) minimax estimation rate (Theorem 1), we can directly compare the limits of estimation in the causal inference setting (established in Theorem 2) with that of the standard nonparametric regression setting. It is well known that the optimal minimax rate for estimating an $\alpha$-regular function is $\Theta(n^{-2\alpha/(2\alpha+d)})$; a classical result due to Stone \cite{stone1982optimal,stone1980optimal}. The result of Theorem 2 (and the tabulated results in Table \ref{tab1}) asserts that the causal effect estimation problem is as hard as the problem of estimating the ``rougher" of the two surfaces $f_0$ and $f_1$ in a standard regression setup. 

The fact that selection bias does not impair the optimal information rate for causal inference is consistent with previous results on minimax-optimal kernel density estimation under selection bias or length bias \cite{efromovich2004density, wu1996minimax,borrajo2017bandwidth,brunel2009nonparametric}. In these settings, selection bias did not affect the optimal minimax rate for density estimation, but the kernel bandwidth optimization strategies that achieve the optimal rate needed to account for selection bias \cite{borrajo2017bandwidth, wu1997cross}. In Section \ref{secadapt}, we show that the same holds for causal inference: in order to achieve the optimal information rate, the strategy for selecting the prior $\Pi$ needs to account for selection bias. This means that even though the optimal information rates in the causal inference and standard regression settings are {\it similar}, the optimal estimation strategies in both setups are {\it different}. 




\section{Rate-adaptive Bayesian Causal Inference}
\label{secadapt}
In Section \ref{secIV}, we have established the optimal rates by which any Bayesian inference procedure can gather information about the causal effect of a treatment from observational data. In this Section, we investigate different strategies for selecting the prior $\Pi$, and study their corresponding achievable information rates. (An optimal prior $\Pi^*$ is one that achieves the optimal information rate $I^*_n$.) A strategy for selecting $\Pi$ comprises the following three modeling choices: 
\begin{enumerate}
\item How to incorporate the treatment assignment variable $\omega$ in the prior $\Pi$?  
\item What function (regularity) class should the prior $\Pi$ place a probability distribution over? 
\item What should be the smoothness (regularity) parameter of the selected function class? 
\end{enumerate}
The~first~modeling~decision~involves~{\it two}~possible~choices. The~{\it first}~choice~is~to~give~no~special~role~to~the~treatment~assignment~indicator~$\omega$,~and~build~a~model~that~treats~it~in~a~manner~similar~to~all~other~features~by~augmenting~it~to~the~feature~space~$\mathcal{X}$. This leads to models of the form 
\[f(x,\omega): \mathcal{X} \times \{0,1\} \to \mathbb{R}.\]
We~refer~to~priors~over~models~of the form above as {\bf Type-I priors}. The {\it second} modeling choice is to let $\omega$ index two different models for the two response surfaces. This leads to models of the form ${\bf f}(x) = [f_0(x),f_1(x)]^T,$ where $f_0 \in \mathcal{F}^{\beta_0}$ and $f_1 \in \mathcal{F}^{\beta_1}$ for some $\beta_0,\, \beta_1 > 0$. We refer to priors over models of the form ${\bf f}(.)$ as {\bf Type-II priors}. 

\begin{figure}[h]
\centering
\includegraphics[width=3.5in]{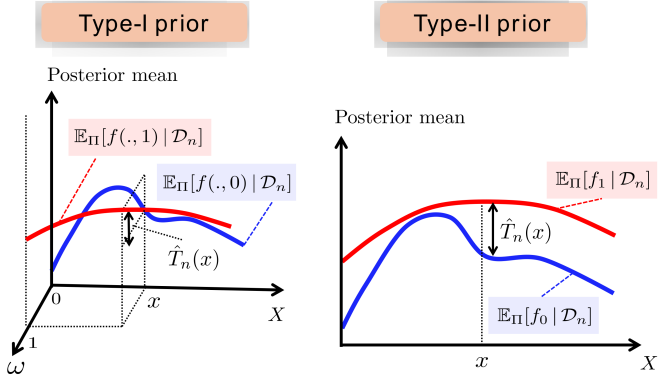}
\caption{\small Depiction for estimates obtained by Type-I and Type-II priors.}
\label{figx}
\end{figure}

Type~I~and~II~priors~induce~different~estimators~for~$T(x)$. The posterior mean estimator for a Type-I prior is given by
\begin{align}   
\hat{T}_n(x) = \mathbb{E}_{\Pi}[f(x,1)\,|\,\mathcal{D}_n]-\mathbb{E}_{\Pi}[f(x,0)\,|\,\mathcal{D}_n], \nonumber
\end{align}
whereas for a Type-II prior, the posterior mean ITE estimator is given by $\hat{T}_n(x) = \mathbb{E}_{\Pi}[{\bf f}^T(x)\,{\bf v}\,|\,\mathcal{D}_n],$ where ${\bf v} = [-1, 1]^T$. Figure \ref{figx} is a pictorial depiction for the posterior mean ITE estimates obtained via Type-I and Type-II priors. 

The main difference between Type-I and Type-II priors is that the former restricts the smoothness of $f(x,\omega)$ on any feature dimension to be the same for $\omega = 0$ and $\omega = 1$. This also entails that the relevant dimensions for the two response surfaces ($\omega = 0$ and $\omega = 1$) need to be the same under a Type-I prior. (This is a direct consequence of the fact that Type-I priors give no special role to the variable $\omega$.) As a result, a priori knowledge (or even data-driven knowledge) on the differences between responses $f_0$ and $f_1$ (e.g. in terms of smoothness levels or relevant dimensions) cannot be incorporated in a Type-I prior. Type-II priors can incorporate such information as they provide separate models for $f_0$ and $f_1$. However, while Type-I priors give a posterior of $f_0$ and $f_1$ using a joint model that is fitted using {\it all} the observational data, Type-II priors use only the data for one population to compute posteriors of one response surface, which can be problematic if the two populations posses highly unbalanced relative sizes (e.g. treated populations are usually much smaller than control populations \cite{foster2011subgroup}).    

In order to better illustrate the difference between Type-I and Type-II priors, we look at their simpler parametric counterparts. A Type-I linear regression model defines $f(x,\omega)$ as a linear function $Y = {\bf \beta}^T\,{\bf x} + \tilde{\gamma}\,\cdot\,\omega + \varepsilon$, where ${\bf \beta} \in \mathbb{R}^d,\, \tilde{\gamma} \in \mathbb{R},$ and $\varepsilon$ is a Gaussian noise variable. (Here the Type-I prior is a prior on the model coefficients ${\bf \beta}$ and $\tilde{\gamma}$) As we can see, this model restricts the two responses $f_0$ and $f_1$ to have the exact same interactions with the features through the coefficients in ${\bf \beta}$. If we know a priori that $f_0$ and $f_1$ have different ``slopes" or different relevant dimensions, we cannot incorporate this knowledge into the model. What would such a model learn? Assuming consistency, the estimated ITE function would be $\hat{T}_n(x) \to \tilde{\gamma}$ everywhere in $\mathcal{X}$. Thus, the restricted nature of a Type-I parametric model led to a constant (non-individualized) estimate of $T(.)$. On the contrary, a Type-II model of the form $Y^{(\omega)} = {\bf \beta}_{\omega}^T\,{\bf x} + \varepsilon,\,\omega \in \{0,1\},$ would allow for learning a linear estimate $\hat{T}_n(x)$ of the true function $T(x)$, with potentially different relevant dimensions for both surfaces. However, Type-II model will only use data with $\omega=w$ to fit the model for $Y^{(w)},\, w \in \{0,1\}$.    

Unlike their parametric counterparts, the nonparametric Type-I and II priors can (in general) learn the ITE function consistently, but how do their information rates compare? Subsection \ref{secadaptA} studies the {\it achievable} information rates for ``oracle" Type-I and Type-II priors that are informed with the true smoothness parameters ($\alpha_0$ and $\alpha_1$) and relevant dimensions of the function classes $\mathcal{F}^{\alpha_0}$ and $\mathcal{F}^{\alpha_1}$. In Subsection \ref{secadaptB}, we study the (more realistic) setting when $\mathcal{F}^{\alpha_0}$ and $\mathcal{F}^{\alpha_1}$ are unknown, and investigate different strategies for adapting the prior $\Pi$ to the smoothness of the treated and control response surface in a data-driven fashion. 

\subsection{Oracle Priors} 
\label{secadaptA}
In~this~Subsection,~we~assume~that~the~true~smoothness~and relevant dimensions for $f_0$ and $f_1$ are known a priori. In the following Theorem, we show that Type-II priors are generally a better modeling choice than Type-I priors.\\
\\
{\bf Theorem~3.~(Sub-optimality~of~Type-I~priors)}~Let~$\Pi_{\beta}^{\circ}$~be~the space of all Type-I priors that give probability one to draws from~$H^{\beta}$,~and~let~$\Pi_{\beta_{0,1}}^{\circ\circ}$~be~the~space~of~all~Type-II~priors~that give probability one to draws from $(H^{\beta_0},H^{\beta_1})$. If $f_0 \in H^{\alpha_0}_{\mathcal{P}_0}$, $f_1 \in H^{\alpha_1}_{\mathcal{P}_1}$, and $\mathcal{P}_0 \neq \mathcal{P}_1$, then
\begin{align}
\inf_{\beta_{0},\beta_{1}}\inf_{\Pi \in \Pi_{\beta_{0,1}}^{\circ\circ}} I_n(\Pi;H^{\alpha_0}_{\mathcal{P}_0},H^{\alpha_1}_{\mathcal{P}_1}) \asymp I^*_n(H^{\alpha_0}_{\mathcal{P}_0},H^{\alpha_1}_{\mathcal{P}_1}),\nonumber\\  
\inf_{\beta}\inf_{\Pi \in \Pi_{\beta}^{\circ}} I_n(\Pi;H^{\alpha_0}_{\mathcal{P}_0},H^{\alpha_1}_{\mathcal{P}_1}) \gtrsim I^*_n(H^{\alpha_0}_{\mathcal{P}_0},H^{\alpha_1}_{\mathcal{P}_1}), \nonumber
\end{align}
where $\gtrsim$ denotes asymptotic inequality.\\
\\ 
{\bf Proof.} See Appendix B. \,\,\, \IEEEQED \\
\\  
Theorem 3 says that if $\mathcal{P}_0 \neq \mathcal{P}_1$, then the information rate that any Type-I prior can achieve is always suboptimal, even if we know the relevant dimensions and the true smoothness of the response surfaces $f_0$ and $f_1$. The Theorem also says that an oracle Type-II prior can achieve the optimal information rate. When the the surfaces $f_0$ and $f_1$ have the same relevant dimensions and the same smoothness, the two priors achieve the same rate. More precisely, the best achievable information rate for a Type-I prior is given by    
\begin{align}
\inf_{\beta}\inf_{\Pi \in \Pi_{\beta}^{\circ}} I_n(\Pi;H^{\alpha_0}_{\mathcal{P}_0},H^{\alpha_1}_{\mathcal{P}_1}) = \Theta\left(n^{\frac{-2(\alpha_0 \wedge \alpha_1)}{2(\alpha_0 \wedge \alpha_1)+|\mathcal{P}_0 \cup \mathcal{P}_1|}}\right),\nonumber
\end{align} 
whereas for Type-II priors, the best achievable rate is 
\begin{align}
\inf_{\beta_{0},\beta_{1}}\inf_{\Pi \in \Pi_{\beta_{0,1}}^{\circ\circ}} I_n(\Pi;H^{\alpha_0}_{\mathcal{P}_0},H^{\alpha_1}_{\mathcal{P}_1}) = \Theta\left(n^{\frac{-2\alpha_0}{2\alpha_0+|\mathcal{P}_0|}} \vee n^{\frac{-2\alpha_1}{2\alpha_1+|\mathcal{P}_1|}}\right).\nonumber
\end{align} 
We note that most state-of-the-art causal inference algorithms, such as causal forests \cite{wager2017estimation}, Bayesian additive regression trees \cite{hill2011bayesian}, and counterfactual regression \cite{johansson2016learning,shalit2016estimating}, use Type-I regression structures for their estimates. The sub-optimality of Type-I priors, highlighted in Theorem 3, suggests that improved estimates can be achieved over state-of-the-art algorithms via a Type-II regression structure.        

We now focus on the second and third modeling questions: on what function space should the prior be placed, and how should we set the regularity of the sample paths drawn from the prior? In the rest of this Section, we assume that the true response surfaces reside in H\"older spaces. One possible prior over H\"older balls is the Gaussian process $\mathcal{GP}(\mbox{Mat\'ern}(\beta))$, with a Mat\'ern covariance kernel and a smoothness parameter $\beta$. (Draws from such a prior are almost surely in a $\beta$-H\"older function space \cite{van2008rates,van2008reproducing}.) In the following Theorem, we characterize the information rates achieved by such a prior.\\
\\ 
{\bf Theorem~4.~(The~Matching~Condition)}~Suppose~that~$f_0$~and $f_1$~are~in~H\"older~spaces~$H^{\alpha_0}$~and~$H^{\alpha_1}$, respectively, and~let
\[\mbox{$\Pi(\beta_0,\beta_1) = (\mathcal{GP}(\mbox{Mat\'ern}(\beta_0)), \mathcal{GP}(\mbox{Mat\'ern}(\beta_1)))$},\]
be a Type-II prior over {\footnotesize $(H^{\beta_0},H^{\beta_1})$}. If {\footnotesize $(\beta_0 \wedge \alpha_0 \wedge \beta_1 \wedge \alpha_1) \geq d/2$}, then we have that
\begin{align}
I_n(\Pi(\beta_0,\beta_1);H^{\alpha_0},H^{\alpha_1}) \lesssim n^{\frac{-2\beta_0}{2\beta_0 + d}} \vee n^{\frac{-2\beta_1}{2\beta_1 + d}},\nonumber
\end{align}    
where posterior consistency holds only if {\footnotesize $\beta_0 \leq \alpha_0$}, and {\footnotesize $\beta_1 \leq \alpha_1$}.\\
\\
{\bf Proof.} See Appendix C. \,\,\, \IEEEQED \\
\\      
For a Type-I prior $\Pi(\beta) = \mathcal{GP}(\mbox{Mat\'ern}(\beta))$, the upper bound on $I_n(\Pi(\beta);H^{\alpha_0},H^{\alpha_1})$ is $n^{\frac{-2\beta}{2\beta + d}}$, with consistency holding for $\beta \leq \alpha$. Using the results of the paper by Castillo in \cite{castillo2008}, the upper bound in Theorem 4 can be shown to be tight. Recall that the optimal information rate for causal inference in H\"older~spaces is $I^*_n(H^{\alpha_0},H^{\alpha_1}) =  n^{\frac{-2(\alpha_0 \wedge \alpha_1)}{2(\alpha_0 \wedge \alpha_1) + d}}$ (Table \ref{tab1}). Theorem 4 quantifies the information rates achieved by a Type-II prior with smoothness levels $\beta_0$ and $\beta_1$. The Theorem says that a prior can achieve the optimal information rate if and only if it {\it captures the smoothness of the rougher of the two response surfaces}. This gives rise to the following {\it matching condition} that a prior $\Pi(\beta_0,\beta_1)$ requires in order to provide an optimal rate:
\begin{align}
\framebox[1.1\width]{\footnotesize $\beta_{\omega} = \alpha_0 \wedge \alpha_1,\, \alpha_{\omega} \leq \beta_{1-\omega} \leq \alpha_{1-\omega}, \omega = \arg \min_{w \in \{0,1\}} \alpha_{w}.$} \nonumber
\end{align}
That is, the regularity of the prior needs to match the rougher of the two surfaces, and the prior over the smoother surface needs to be at least as smooth as the rougher surface. Consistency holds only if the prior is at least as smooth as the true response, since otherwise the response surfaces would not be contained in the support of the prior. Note that Theorem 4 assumes that the true response surfaces exhibit a H\"older-type regularity, and that the prior $\Pi(\beta_0,\beta_1)$ is placed on a reproducing kernel Hilbert space with a particular kernel structure. While proving that the matching condition holds for general priors and function spaces is technically daunting, we believe that (given the results in Table I) the matching condition in Theorem 4 would hold for other notions of regularity (e.g. Sobolev, Lipschitz, etc), and for a wide range of practical priors. For instance, Theorem 4 holds for Gaussian processes with re-scaled squared exponential kernels \cite{pati2015optimal}.

To sum up this Subsection, we summarize the conclusions distilled from our analyses of the achievable rates for oracle priors. Priors of Type II are generally a better design choice compared to priors of Type I, especially when the two response surfaces exhibit different forms of heterogeneity. In order to achieve the optimal information rate, a typical condition is that the regularity of the prior needs to match that of the rougher of the two response surfaces. Since in practice we (generally) do not know the true smoothness of the response surfaces, we cannot build a prior that satisfies the matching condition. Practical causal inference thus requires {\it adapting} the prior to the smoothness of the true function in a data-driven fashion; we discuss this in the next Subsection.   
   
\subsection{Rate-adaptive Data-driven Priors}
\label{secadaptB}
Note that, unlike in standard nonparametric regression, adapting the regularity of the prior for the causal inference inference task entails a mixed problem of testing and estimation, i.e. we need to test whether $\alpha_0$ is less than $\alpha_1$, and then estimate $\alpha_0$ (or $\alpha_0$). Hence, one would expect that the prior adaptation methods used in standard regression problems would not necessarily suffice in the causal inference setup. Prior adaptation can be implemented via {\it hierarchical Bayes} or {\it empirical Bayes} methods. Hierarchical Bayes methods specify a prior over $\beta = (\beta_0,\beta_1)$ (also known as the {\it hyper-prior} \cite{sniekers2015adaptive}), and then obtain a posterior over the regularity parameters in a fully Bayesian fashion. Empirical Bayes simply obtains a point estimate $\hat{\beta}_n$ of $\beta$, and then conducts inference via the prior specified by $\hat{\beta}_n$. We focus on empirical Bayes methods since the hierarchical methods are often impractically expensive in terms of memory and computational requirements. A prior $\Pi_{\hat{\beta}_n}$ induced by $\hat{\beta}_n$ (obtained via empirical Bayes) is called {\it rate-adaptive} if it achieves the optimal information rate, i.e. $I_n(\Pi_{\hat{\beta}_n}) = I^*_n$.\\  

In the rest of this Subsection, we show that marginal likelihood maximization, which is the dominant strategy for empirical Bayes adaptation in standard nonparametric regression \cite{hoffmann2015adaptive,sniekers2015adaptive}, can fail to adapt to $\alpha_0 \wedge \alpha_1$ in the general case when $\alpha_0 \neq \alpha_1$. (This is crucial since in most practical problems of interest, the treated and control response surfaces have different levels of heterogeneity \cite{xie2013population}.) We then propose a novel {\it information-based empirical Bayes} strategy, and prove that it asymptotically satisfies the matching condition in Theorem 4. Finally, we conclude the Subsection by identifying candidate function spaces over which we can define the prior $\Pi$ such that we are able to both adapt to functions in H\"older spaces, and also conduct practical Bayesian inference in an algorithmically efficient manner.\\ 

\subsubsection{Information-based Empirical Bayes}
\begin{figure}[t]
\centering
    \begin{subfigure}[h]{0.45\textwidth}
        \includegraphics[width=\textwidth]{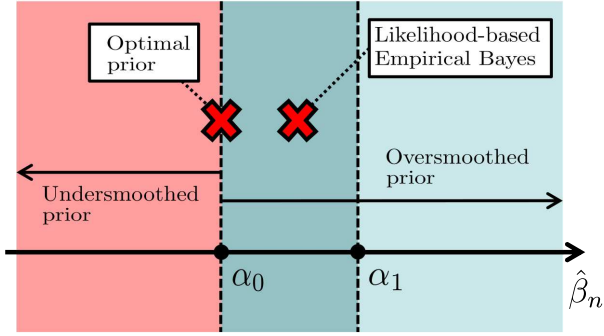}
        \caption{An exemplary data-driven prior obtained via the likelihood-based empirical Bayes method.}
        \label{figy}
    \end{subfigure}
        \begin{subfigure}[h]{0.45\textwidth}
        \includegraphics[width=\textwidth]{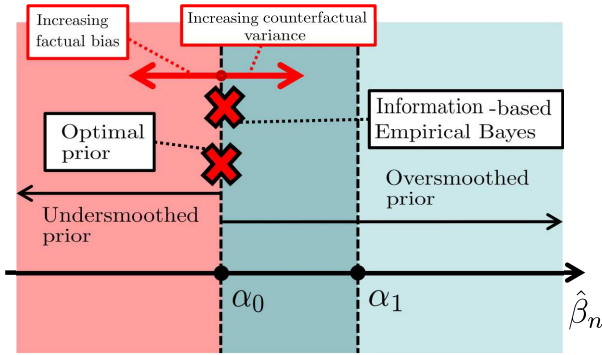}
        \caption{An illustration for the factual bias and counterfactual variance trade-off.}
        \label{fig:fig2}
    \end{subfigure}
\caption{\small Pictorial depiction for the operation of likelihood-based and information-based empirical Bayes adaptation methods.}
\label{fgggg}
\end{figure}
To see why the marginal likelihood-based empirical Bayes method may fail in adapting priors for causal inference, consider the following example. Suppose that $f_0 \in H^{\alpha_0}$ and $f_1 \in H^{\alpha_1},$ where $\alpha_0 < \alpha_1$. Let $\Pi(\hat{\beta}_n)$ be a Type-I data-driven prior, where $\hat{\beta}_n$ is an empirical Bayes estimate of $\alpha_0 \wedge \alpha_1$. For the likelihood-based empirical Bayes, $\hat{\beta}_n$ is obtained by maximizing the marginal likelihood $d\mathbb{P}(\mathcal{D}_n\,|\,\beta)$ with respect to $\beta$. Note that since $f_0$ and $f_1$ possess different smoothness parameters, then the ``true" model for generating $\mathcal{D}_n$ is characterized by a likelihood function $d\mathbb{P}(\mathcal{D}_n\,|\,\alpha_0, \alpha_1)$. Assume that the true model $d\mathbb{P}(\mathcal{D}_n\,|\,\alpha_0, \alpha_1)$ is {\it identifiable}, i.e. the mapping $(\alpha_0, \alpha_1) \mapsto \mathbb{P}$ is one-to-one. Type-I priors re-parametrize the observation model so that the likelihood function $d\mathbb{P}(\mathcal{D}_n\,|\,\beta)$ is parametrized with a single smoothness parameter $\beta$. Hence, as long as $\alpha_0 \neq \alpha_1$, the new parametrization renders an unidentifiable model, since the mapping $\beta \mapsto \mathbb{P}$ is not one-to-one (i.e. different combinations of $\alpha_0$ and $\alpha_1$ can map to the same $\beta$). This means that in this case likelihood-based empirical Bayes would never satisfy the matching condition in Theorem 4, even in the limit of infinite samples ($n \uparrow \infty$). In most practical Bayesian models (e.g. Gaussian processes), the empirical Bayes estimate $\hat{\beta}_n$ will be in the interval $(\alpha_0,\alpha_1)$ with high probability as depicted in Figure \ref{figy}. This means that with high probability, the likelihood-based empirical Bayes method will prompt an {\it oversmoothed} prior, from which all draws are smoother than the true ITE function, leading to a suboptimal information rate. 

The failure of likelihood-based empirical Bayes in the causal inference setup is not surprising as maximum likelihood adaptation is only optimal in the sense of minimizing the Kullback-Leibler loss for the individual potential outcomes. {\it Optimal prior adaptation in our setup should be tailored to the causal inference task}. Hence, we propose an information-based empirical Bayes scheme in which, instead of maximizing the marginal likelihood, we pick the smoothness level $\hat{\beta}_n$ that minimizes the posterior Bayesian KL divergence, i.e.   
\begin{align}
\hat{\beta}_n &= \arg \min_{\beta} \mathbb{E}_{\Pi(.\,|\,\mathcal{D}_n,\beta)}[\,\mathbb{D}_n(\Pi(\beta); f_0, f_1)\,|\,\mathcal{D}_n\,] \nonumber \\
&= \arg \min_{\beta} \mathbb{E}_{\Pi(.\,|\,\mathcal{D}_n,\beta)}[\,\mathbb{E}_{x}[\,D_{\mathrm {\small KL}}(P(x)\,\|\,Q_{\mathcal{D}_n}(x))\,]\,]. 
\label{eqBay}
\end{align} 
The information-based empirical Bayes estimator is simply a Bayesian estimator of $\beta$ with the loss function being the posterior KL risk in (\ref{eq4}). Unlike the likelihood-based method, the objective in (\ref{eqBay}) is an direct measure for the quality of causal inference conducted with a prior $\Pi_{\beta}$. In the following Theorem, we show that $\hat{\beta}_n$ asymptotically satisfies the matching condition in Theorem 4.\\
\\
{\bf Theorem~5.~(Asymptotic~Matching)}~Suppose~that~$f_0$~and~$f_1$ belong~to~the~H\"older~spaces~$H^{\alpha_0}$~and~$H^{\alpha_1}$,~respectively, and let $\Pi(\beta)$ be a prior over H\"older~space with order $\beta$. If $\hat{\beta}_n$ is obtained as in (\ref{eqBay}) using cross-validation, then under certain regularity conditions we have that $\hat{\beta}_n \overset{p}{\to} (\alpha_{0} \wedge \alpha_{1})$.\\  
\\
{\bf Proof.} See Appendix D. \,\,\, \IEEEQED \\
\\    
Theorem~5~says~that~the~information-based~empirical~Bayes estimator is consistent. That is, the estimate $\hat{\beta}_n$ will eventually converge to $\alpha_{0} \wedge \alpha_{1}$ as $n \to\infty$. Note that this is a weaker result than adaptivity: consistency of $\hat{\beta}_n$ does not imply that the corresponding prior will necessarily achieve the optimal information rate. However, the consistency result in Theorem 5 is both strongly suggestive of adaptivity, and also indicative of the superiority of the information-based empirical Bayes method to the likelihood-based approach. 

Note that, while the information-based empirical Bayes approach guarantees the asymptotic recovery of $\alpha_{0} \wedge \alpha_{1}$, it can still undersmooth the prior for the smoother response surface. This can problematic if we wish the posterior credible interval on $T(x)$ to be ``honest", i.e. possess frequentist coverage \cite{knapik2011bayesian,wager2017estimation,athey2016recursive}. A more flexible Type-II prior that assigns different smoothness parameters $\beta_0$ and $\beta_1$ to response surfaces $f_0$ and $f_1$ can potentially guarantee honest frequentist coverage in a manner similar to that provided by causal forests \cite{wager2017estimation}. As a consequence of Theorem 1, it turns out that the information-based empirical Bayes estimator in (\ref{eqBay}) is structurally similar to the risk-based empirical Bayes adaptation method proposed in \cite{sniekers2015adaptive}. Hence, we conjecture that our proposed empirical Bayes procedure can guarantee frequentist coverage for the estimated causal effects under some conditions \cite{sniekers2015adaptive}.\\

\subsubsection{Concrete Priors for Bayesian Causal Inference} Assuming that $f_0$ and $f_1$ belong to H\"older spaces, what concrete priors should one use in order to achieve the optimal information rates? We have already shown (in Theorem 4) that the Gaussian process prior $\Pi(\beta) = \mathcal{GP}(\mbox{Mat\'ern}(\beta))$, which places a probability distribution over a H\"older space with regularity $\beta$, can achieve the optimal information rate under the matching condition. Gaussian processes, in general, place a probability distribution on a reproducing kernel Hilbert space (RKHS) \cite{rasmussen2006gaussian,alvarez2012kernels}, the nature of which is determined by the kernel structure. It is worth mentioning that for kernels other than the Mat\'ern kernel, the optimal rate might not be achievable. For instance, using the squared exponential kernel would lead to a suboptimal information rate of $(\log(n))^{-(\alpha_0 \wedge \alpha_1)/2+d/4}$, whereas spline kernels achieved a rate of $(n/\log(n))^{-2(\alpha_0 \wedge \alpha_1)/(2(\alpha_0 \wedge \alpha_1) + d)},$ which is optimal up to a logarithmic factor \cite{van2008reproducing}. In order for such kernels to achieve the optimal rates, their smoothness parameters (e.g. the length-scale parameter of the radial basis kernel) need to be re-scaled with the size of the observational data as explained in \cite{pati2015optimal}. Selection of the right kernel can be based either on prior knowledge of the response surfaces, or through a model selection procedure based on the information-theoretic criterion in (\ref{eqBay}).   

Another possible option for Bayesian nonparametric priors place their probability mass on the space of piece-wise constant functions (trees) \cite{hahn2017bayesian,hahn2017bayesian,hill2011bayesian}. The machine learning object operating on those spaces is the {\it Bayesian additive regression trees} (BART) algorithm, which was especially proven successful in causal inference problems, and was one of the winning algorithms in the 2016 Atlantic Causal Inference Conference Competition\footnote{http://jenniferhill7.wixsite.com/acic-2016}. Since BART places a prior on a space of non-differentiable (piece-wise constant) functions, one would expect that the information rates achieved by BART would be inferior to those achieved by a GP. A carefully designed BART can only achieve a near-optimal rate of $(n/\log(n))^{-2(\alpha_0 \wedge \alpha_1)/(2\alpha_0 \wedge \alpha_1 + d)}$ \cite{linero2017bayesian,rockova2017posterior,hahn2017bayesian}. Our conclusion is that a Gaussian process is a better choice for causal modeling, not only because it can achieve better rates than BART, but also because its relatively tractable nature would allow for an easy implementation for the information-based empirical Bayes scheme in (\ref{eqBay}).  

Finally, we note that if we know a priori which response surface is rougher, then prior adaptation can be achieved very easily by tuning the prior smoothness to the population that correspond to the rougher surface only. Such an adaptation can be done through the conventional marginal likelihood maximization method. It is worth mentioning though that while practitioners may know which surface is rougher a priori, it is less likely that in a high-dimensional space we would know ahead of time which variables are relevant to which surface. As we can see in the discussion after Theorem 3, the information rate is bottle-necked by complexity and not just smoothness. A smoother surface with more relevant dimensions can still bottle-neck the information rate. So practitioners should consider variable selection, and not just smoothness estimation, as a means to adapt the prior.
 
\section{Practical Rate-adaptive Causal Inference with Multitask Gaussian Process Priors}
\label{mtgp}
The previous Section provided a detailed recipe for the informationally optimal Bayesian causal inference procedure. In particular, inference should be conducted through a Type-II Gaussian process prior on an RKHS space (Theorem 3 and Subsection \ref{secadaptB}). Moreover, the RKHS space should be defined through a Mat\'ern covariance kernel with parameters $\beta_0$ and $\beta_1$ for response surfaces $f_0$ and $f_1$ (Subsection \ref{secadaptB}), and the parameters ${\bf \beta} = (\beta_0,\beta_1)$ should be optimized via the information-based empirical Bayes procedure in (\ref{eqBay}). In this Section, we construct a practical learning algorithm that follows this recipe.

Type-II GP priors place a probability distribution on functions ${\bf f}: \mathcal{X} \to \mathbb{R}^2$ in a {\it vector-valued Reproducing Kernel Hilbert Space} (vvRKHS). A vvRKHS $\mathcal{H}_{{\bf K}}$ is equipped with an inner product $\langle .,. \rangle_{\mathcal{H}_{{\bf K}}}$, and a {\it reproducing kernel} ${\bf K}: \mathcal{X} \times \mathcal{X} \rightarrow \mathbb{R}^{2 \times 2}$, where ${\bf K}$ is a (symmetric) positive semi-definite matrix-valued function \cite{bonilla2008multi,alvarez2012kernels,rasmussen2006gaussian,alaa2017bayesian}. Note that by operating in a vvRKHS we get the algorithmic advantage of being able to conduct posterior inference in an infinite-dimensional function space by estimating a finite number of coefficients evaluated at the input feature points (this is a consequence of the well-known {\it representer Theorem} \cite{scholkopf2001generalized}). GP regression in vvRKHS is often associated with {\it multi-task learning} \cite{alvarez2012kernels}, and the corresponding GP is often known as a {\it multi-task} GP \cite{bonilla2008multi}. Multi-task learning is a common setup in machine learning where one model shares parameters between different tasks to improve statistical efficiency. The results of Theorem 3 can be thought of as suggesting multi-task learning as a framework for causal inference, where learning each of the potential outcomes~($f_0$~and~$f_1$)~is~thought~of~as~a~separate~learning~task, and a single model is used to execute the two tasks simultaneously. 

We chose the Mat\'ern covariance kernel as the underlying regularity of the vvRKHS since it can achieve the optimal information rate (see Appendix E). In order to avoid undersmoothing any of the two surfaces, we also chose to assign separate smoothness parameters $\beta_0$ and $\beta_1$ to $f_0$ and $f_1$, respectively. Standard {\it intrinsic coregionalization models} for vector-valued kernels impose the same covariance parameters for all outputs \cite{bonilla2008multi}, which implies that the prior will have the same smoothness on both $f_0$ and $f_1$. Thus, we constructed a {\it linear model of coregionalization} (LMC) \cite{alvarez2012kernels}, which mixes two intrinsic coregionalization models as follows 
\begin{align}
{\bf K}_\theta(x, x^{\prime}) = {\bf A}\,k_{0}(x, x^{\prime}) + {\bf B}\, k_{1}(x, x^{\prime}),\nonumber
\end{align}
where $k_{\omega}(x, x^{\prime}) = \mbox{Mat\'ern}(\beta_\omega),\, \omega \in \{0,1\},$ whereas ${\bf A}$ and ${\bf B}$ are given by 
\begin{align}
{\bf A} = \begin{bmatrix}
    a^2_{00} & a_{01} \\
    a_{10} & \epsilon 
\end{bmatrix},\,\, 
{\bf B} = \begin{bmatrix}
    \epsilon & b_{01} \\
    b_{10} & b_{11} 
\end{bmatrix},
\label{eqKern}
\end{align}
where $\epsilon \to 0$ is a small positive number that is determined a priori and kept fixed during the prior adaptation procedure. The LCM kernel structure in (\ref{eqKern}) ensures that the response surfaces $f_0$ and $f_1$ have smoothness levels $\beta_0$ abd $\beta_1$ respectively. The constant $\epsilon$ ensures that ${\bf K}_\theta(x, x^{\prime})$ is positive semi-definite for any selection of the other parameters. The parameters $a_{00}$ and $b_{11}$ represent the variances of $f_0$ and $f_1$, whereas all other variables $(a_{01},a_{10},b_{01},b_{10})$ are correlation variables that share information among the two learning tasks (learning $f_0$ and $f_1$). The set of {\it all} kernel parameters is denoted as ${\bf \beta}$. Given a set of ``hyper-parameters" ${\bf \beta}$, the ITE function estimate $\hat{T}_n$ is obtained in terms of the posterior mean\footnote{Closed-form expressions for the posterior mean of a multi-task GP can be found in \cite{rasmussen2006gaussian,bonilla2008multi}.} as follows: $\hat{T}_n = \mathbb{E}_{\Pi_{\bf \beta}}[\,{\bf f}^T{\bf v}\,|\,\mathcal{D}_n],$ where ${\bf v} = [-1,1]^T$.   

Now that we completely specified the multi-task GP prior for a given hyper-parameter set ${\bf \beta}$, the only remaining ingredient in the recipe is to implement the information-based empirical Bayes adaptation criterion in (\ref{eqBay}). The following Theorem gives an insightful decomposition of the information-based empirical Bayes objective for the multi-task GP model. (In the following Theorem, ${\bf Y^{(W)}} = [Y_i^{(\omega_i)}]_i$ and ${\bf Y^{(1-W)}} = [Y_i^{(1-\omega_i)}]_i$ are vectors comprising all factual and counterfactual outcomes associated with an observational dataset $\mathcal{D}_n$.)\\
\\
{\bf Theorem~6.~(Factual~bias~and~counterfactual~variance~decomposition)} The minimizer ${\bf \beta}^{*}$ of the information-based empirical Bayes adaptation criterion in (\ref{eqBay}) is given by
\begin{align}
\arg \, \min_{\beta} \underbrace{\left\|{\bf Y^{(W)}}-\mathbb{E}_{\Pi_{\beta}}[\,{\bf f}\,|\,\mathcal{D}_n\,]\right\|_2^2}_{\mbox{\small Factual bias}} + \underbrace{\left\|\mbox{Var}_{\Pi_{\beta}}[\,{\bf Y^{(1-W)}}\,|\,\mathcal{D}_n\,]\right\|_1}_{\mbox{\small Counterfactual variance}},\nonumber
\end{align}
where $\mbox{Var}_{\Pi_{\beta}}$ is the posterior variance and $\|.\|_{p}$ is the $p$-norm.\\  
{\bf Proof.} See Appendix E. \,\,\, \IEEEQED \\
\\    
Theorem 6 states that, when the prior is specified as a multi-task GP, the information-based empirical Bayes criterion in (\ref{eqBay}) decomposes to {\it factual bias} and {\it counterfactual variance} terms\footnote{The objective function in Theorem 6 can be easily optimized via a leave-one-out cross-validation procedure. Refer to \cite{alaa2017bayesian} for a detailed explanation.}. The factual bias term quantifies the empirical error in the observed factual outcome that results from selecting a particular smoothness level $\beta$. In that sense, the factual bias is a measure of the goodness-of-fit for the posterior mean resulting from a prior smoothness $\beta$. On the other hand, the counterfactual variance term quantifies the posterior uncertainty that would be induced in the unobserved counterfactual outcomes when selecting a smoothness level $\beta$. A small value for $\beta$ would lead to a rough posterior mean function, which corresponds to a good empirical fit for the data. On the contrary, a small value for $\beta$ would induce large uncertainty in the unobserved outcomes, which corresponds to large uncertainty in the counterfactual outcomes. The couterfactual variance thus acts as a regularizer for the factual bias that helps solving the joint testing-estimation problem of identifying the minimum of $\alpha_0$ and $\alpha_1$, and estimating the value of  $\alpha_0 \wedge \alpha_1$. That is, the regularizer attempts to protect the prior from falsely recognizing either $\alpha_0$ or $\alpha_1$ as being very low just because it over-fit the factual outcomes, and hence underestimating the true $\alpha_0 \wedge \alpha_1$, thereby undersmoothing the prior and giving rise to a suboptimal information rate. The two terms work in opposite directions as shown in Figure \ref{fig:fig2}: factual bias pushes for undersmoothed priors and counterfactual variance pushes for oversmoothed priors. Theorem 6 says that the resulting prior will lie on the optimal boundary in the large data limit.

Finally, we note that the factual bias and counterfactual variance trade-off automatically handles selection bias. That is, when there is a poor overlap between the treated and control populations, the posterior counterfactual variances would tend to be higher, and the information-based empirical Bayes method would tend to oversmooth the prior rather than fitting the factual data. {\it Selection bias does not affect the optimal information rate, but it does affect the optimal strategy for achieving that rate as long as we decide to share parameters and data points between our models for the potential outcomes.}   

  
   


\section{Experiments}
\label{expsec}
We sought to evaluate the finite-sample performance of the Bayesian causal inference procedure proposed in Section \ref{mtgp}, and compare it with state-of-the-art causal inference models. Causal inference models are hard to evaluate \cite{tran2016model}, and obviously, it is impossible to validate a causal model using real-world data due to the absence of counterfactual outcomes. A common approach for evaluating causal models, which we follow in this paper, is to validate the model's predictions/estimates in a semi-synthetic dataset for which artificial counterfactual outcomes are randomly generated via a predefined probabilistic model. To ensure a fair and objective comparison, we did not design the semi-synthetic dataset used in the experiments by ourselves, but rather used the (standard) semi-synthetic experimental setup designed by Hill in \cite{hill2011bayesian}. In this setup, the features and treatment assignments are real but outcomes are simulated. The experimental setup was based on the IHDP dataset, a public dataset for data from a randomized clinical trial. We describe the dataset in more detail in the following Subsection.

\subsection{The IHDP dataset}
The Infant Health and Development Program (IHDP) is an interventional program that is intended to enhance the cognitive and health status of low birth weight, premature infants through pediatric follow-ups and parent support groups \cite{hill2011bayesian}. The semi-simulated dataset in \cite{hill2011bayesian,johansson2016learning,shalit2016estimating} is based on features for premature infants enrolled in a real randomized experiment that evaluated the impact of the IHDP on the subjects' IQ scores at the age of three. Because the data was originally collected from a randomized trial, selection bias was introduced in the treatment assignment variable by removing a subset of the treated population. All outcomes (response surfaces) are simulated. The response surface data generation process was not designed to favor our method: we used the standard non-linear "Response Surface B" setting in \cite{hill2011bayesian}. The dataset comprises 747 subjects (608 control and 139 treated), and there are 25 features associated with each subject.

\begin{table*}[h]
\centering 
\caption{\small \textsc{Simulation results for the IHDP dataset. Numerical values correspond to the average PEHE \, $\pm$\, 95$\%$ confidence intervals.}}
\begin{tabular}{cccccccccc} \toprule 
		   &  & & {\small In-sample} & {\small Out-of-sample} &  & & {\small In-sample} & {\small Out-of-sample}  \\
			  & & & {\small $\sqrt{\mbox{PEHE}}$} & {\small $\sqrt{\mbox{PEHE}}$} & & & {\small $\sqrt{\mbox{PEHE}}$} & {\small $\sqrt{\mbox{PEHE}}$} \\\midrule
    \textcolor{blue}{$\heartsuit$} & {\bf MTGP (Type-II)} & & {\bf 0.59\, $\pm$\, 0.01} & {\bf 0.76 \, $\pm$\, 0.01} & \textcolor{blue}{$\diamondsuit$} &  Causal MARS & 1.66 \, $\pm$\, 0.10 & 1.74 \, $\pm$\, 0.10 \\
       & {\bf GP (Type-I)} & & 1.85\, $\pm$\, 0.12 & 2.10 \, $\pm$\, 0.16 & \textcolor{blue}{$\square$} & NN-1 & 3.56 \, $\pm$\, 0.20 &  3.64 \, $\pm$\, 0.20  \\ 
		\textcolor{blue}{$\clubsuit$} & BART & & 2.0\, $\pm$\, 0.13 & 2.2\, $\pm$\, 0.15 & & AdaBoost-1 & 4.53 \, $\pm$\, 0.31 & 4.56 \, $\pm$\, 0.31 \\
		   & CF & & 2.4\, $\pm$\, 0.14 & 2.8\, $\pm$\, 0.18 & & XGBoost-1 & 2.97 \, $\pm$\, 0.21 & 3.04 \, $\pm$\, 0.21 \\
			 & RF-1 & & 2.7\, $\pm$\, 0.24 & 2.9\, $\pm$\, 0.25 & & LR-1 & 5.06 \, $\pm$\, 0.35 & 5.05 \, $\pm$\, 0.35 \\
       & RF-2 & & 1.4\, $\pm$\, 0.07 & 2.2\, $\pm$\, 0.16 & \textcolor{blue}{$\otimes$} & NN-2 & 3.36 \, $\pm$\, 0.13 & 3.46 \, $\pm$\ 0.14 \\ 	
		\textcolor{blue}{$\spadesuit$} & BNN & & 2.1\, $\pm$\, 0.11 & 2.2\, $\pm$\, 0.13 & & AdaBoost-2 & 2.40 \, $\pm$\, 0.17 & 2.79 \, $\pm$\, 0.20 \\
       & CFRW & & 1.0\, $\pm$\, 0.07 & 1.2\, $\pm$\, 0.08 & & XGBoost-2 & 1.46 \, $\pm$\, 0.08 & 1.98 \, $\pm$\, 0.15 \\ 
		\textcolor{blue}{$\bigstar$} & $k$NN & & 2.69 \, $\pm$\, 0.17 & 4.0\, $\pm$\, 0.21 & & LR-2 & 1.85 \, $\pm$\, 0.10
 & 1.94 \, $\pm$\, 0.12 \\ 
       & PSM & & 4.9\, $\pm$\, 0.31 & 4.9\, $\pm$\, 0.31 & \textcolor{blue}{$\odot$} & TMLE & 5.27 \, $\pm$\, 0.35 & 5.27 \, $\pm$\, 0.35  \\ \bottomrule
\end{tabular}
\label{TabRes1} 
\end{table*} 

\subsection{Benchmarks} 
We compared our algorithm with various causal models and standard machine learning benchmarks which we list in what follows: \textcolor{blue}{$\clubsuit$ Tree-based methods} (BART \cite{hill2011bayesian,hahn2017bayesian,rockova2017posterior}, causal forests (CF) \cite{wager2017estimation,athey2016recursive}, \textcolor{blue}{$\spadesuit$ Balancing counterfactual regression} (balancing neural networks (BNN) \cite{johansson2016learning}, and counterfactual regression with Wasserstein distance metric (CFRW) \cite{shalit2016estimating}), \textcolor{blue}{$\bigstar$ Propensity-based and matching methods} ($k$ nearest-neighbor ($k$NN), propensity score matching (PSM)), a \textcolor{blue}{$\diamondsuit$ nonparametric spline regression} model (causal MARS \cite{powers2017some}), and \textcolor{blue}{$\odot$ Doubly-robust methods} (Targeted maximum likelihood (TML) \cite{porter2011relative}). We also compared the performance of our model with standard machine learning benchmarks, including linear regression (LR), random forests (RF), AdaBoost, XGBoost, and neural networks (NN). We evaluated two different variants of all the machine learning benchmarks: a \textcolor{blue}{$\square$ Type-I regression structure}, in which we use the treatment assignment variable as an input feature to the machine leaning algorithm, and a \textcolor{blue}{$\otimes$ Type-II regression structure}, in which we fit two separate models for treated and control populations. We compare all these benchmarks with our proposed model: a Type-II multi-task GP prior (MTGP) with a Mat\'ern kernel optimized through information-based empirical Bayes. We also compare the proposed model with a Type-I multi-task GP model (with a Mat\'ern kernel) optimized through likelihood-based empirical Bayes in order to verify the conclusions drawn from our analyses. 

All machine learning benchmarks had their hyperparameters optimized via grid search using a held-out validation set. Hyper-parameter optimization was using the mean square error in the observed factual outcomes as the optimization objective. For BART, we used the default prior as in \cite{hill2011bayesian}, and did not tune the model's hyper-parameters. For BNN and CFRW, we used the neural network configurations reported in \cite{johansson2016learning} and \cite{shalit2016estimating}. Causal MARS was implemented as described in \cite{powers2017some}. PSM was implemented as described in \cite{hill2011bayesian}, and its performance was obtained by assuming that every patient's estimated ITE is equal to the average treatment effect estimated by PSM. All benchmarks were implemented in \texttt{Python}, with the exception of BART, causal forests and TMLE, all of which were implemented in \texttt{R}. We used the \texttt{R} libraries \texttt{bartMachine}, \texttt{grf}, and \texttt{tmle} for the implementation of BART, causal forests and TMLE, respectively. Our method was implemented in \texttt{Python} using \texttt{GPy}, a library for Gaussian processes \cite{gpy2014}.      

\subsection{Evaluation} 
We evaluate the performance of all benchmarks by reporting the square-root of the PEHE. The empirical PEHE is estimated as $\mbox{PEHE} = \frac{1}{n}\sum_{i=1}^{n}((f_1(X_i)-f_0(X_i))-\mathbb{E}[Y^{(1)}_i-Y^{(0)}_i|X_i=x])^2$, where $f_1(X_i)-f_0(X_i)$ is the estimated treatment effect. We evaluate the PEHE of all algorithms via a Monte Carlo simulation with 1000 realizations of the IHDP semi-synthetic model, where in each experiment/realization we run all the benchmarks with a 60/20/20 train-validation-test splits. (For models that do not need hyper-parameter tuning, such as BART and our GP models, the entire training set is used to compute the posterior distributions.) We report both the in-sample and out-of-sample PEHE estimates: the former corresponds to the accuracy of the estimated ITE in a retrospective cohort study, whereas the latter corresponds to the performance of a clinical decision support system that provides out-of-sample patients with ITE estimates \cite{shalit2016estimating}. The in-sample PEHE results are non-trivial since we never observe counterfactuals even in the training phase. Recall that, from Theorem 1, we know that the achieved information rate by a Bayesian inference procedure is equivalent to the PEHE estimation rate. Thus, the PEHE performance is a direct proxy of the achieved information rate, and since it is an essentially frequentist quantity, we can use it to compare the performance of our model with the frequentist benchmarks.         

\subsection{Results} 
As can be seen in Table \ref{TabRes1}, the proposed Bayesian inference algorithm (Type-II MTGP) outperforms all other benchmarks in terms of the (in-sample and out-of-sample) PEHE. This result suggests that the proposed model was capable of adapting its prior to the data, and may have achieved the optimal (or a near-optimal) information rate. The PEHE results in Table \ref{TabRes1} are the averages of 1000 experiments with 1000 different random realizations of the semi-synthetic outcome model. This means that our algorithm is consistently outperforming all other benchmarks as it is displaying a very tight confidence interval. 

The benefit of the information-based empirical Bayes method manifests in the comparison with the Type-I MTGP prior optimized via likelihood-based empirical Bayes. The performance gain of the Type-II MTGP prior with respect to the Type-I MTGP prior results from the fact that the two response surfaces in the synthetic outcomes model have different levels of heterogeneity (the control response is non-linear whereas the treated response is linear. See the description of Response surface B in \cite{hill2011bayesian}). Our algorithm is also performing better than all other nonparametric tree-based algorithms. This is expected since, as we have discussed earlier in Subsection \ref{secadaptB}, an oracle BART prior can only achieve the optimal information rate up to a logarithmic factor. With the default prior, it is expected that BART would display a slow information rate as compared to our adapted, information-optimal Mat\'ern kernel prior. Similar insights apply to the frequentist random forest algorithms, which approximates the true regression functions through non-differentiable, piecewise functions (trees), and hence is inevitably suboptimal in terms of the achievable minimax estimation rate.

Our model also outperforms all the standard machine learning benchmarks, whether the ones trained with a Type-I regression structure, or those trained with a Type-II structure. We believe that this is because our model outperforms the standard machine learning benchmarks since the information-based empirical Bayes method provides a natural protection against selection bias (via the counterfactual variance regularization). Selection bias introduces a mismatch between the training and testing datasets for all the machine learning benchmarks (i.e. a covariate shift \cite{johansson2016learning}), and hence all machine learning methods exhibit high generalization errors. 

\section{Conclusions}
In this paper, we studied the problem of estimating the causal effect of an intervention on {\it individual} subjects using observational data in the Bayesian nonparametric framework. We characterized the optimal Kullback-Leibler information rate that can be achieved by any learning procedure, and showed that it depends on the dimensionality of the feature space, and the smoothness of the ``rougher" of the two potential outcomes. We characterized the priors that are capable of achieving the optimal information rates, and proposed a novel empirical Bayes procedure that is adapts the Bayesian prior to the causal effect function through an information-theoretic criterion. Finally, we used the conclusions drawn from our analysis and designed a practical Bayesian causal inference algorithm with a multi-task Gaussian process, and showed that it significantly outperforms the state-of-the-art causal inference models through experiments conducted on a standard semi-synthetic dataset.

\appendices
\section{Proof of Theorem 2}
\renewcommand{\theequation}{\thesection.\arabic{equation}}
We start by establishing an asymptotic equivalence between the KL risk and the frequentist loss in the $L_2(\mathbb{P})$ norm, i.e. 
\begin{align}
\mathbb{D}_n(\Pi; f_0, f_1,\gamma) \asymp \mathbb{E}_{\mathcal{D}_n}\left[\,\big\Vert\mathbb{E}_{\Pi}\left[\,T\,|\,\mathcal{D}_n\,\right]-T\big\Vert^2_2\,\right].
\label{AAAA0}
\end{align}
Note that, from Theorem 1, we already know that since the expected KL risk is bounded above by the $L_2(\mathbb{P})$ loss (with a constant factor), then it follows that: 
\begin{align}
\mathbb{D}_n(\Pi; f_0, f_1,\gamma) \lesssim \mathbb{E}_{\mathcal{D}_n}\left[\,\big\Vert\mathbb{E}_{\Pi}\left[\,T\,|\,\mathcal{D}_n\,\right]-T\big\Vert^2_2\,\right].
\label{AAAA1}
\end{align}
Recall from (\ref{eq4}) that the KL risk is given by
\begin{align} 
\mathbb{D}_n(\Pi; f_0, f_1,\gamma) = \mathbb{E}_{x}\left[\,\mathbb{E}_{\mathcal{D}_n}\left[D_{\mathrm {\small KL}}\left(P(x)\,\left\|\,Q_{\mathcal{D}_n}(x)\right)\right.\right]\,\right]. \nonumber
\end{align}
Using Pinsker's inequality \cite[Lemma 11.6.1]{cover2012elements}, the KL divergence can be bounded below as follows:
\begin{align} 
\|P(x)-Q_{\mathcal{D}_n}(x)\|_{TV} \leq \sqrt{\frac{1}{2} D_{\mathrm {\small KL}}\left(P(x)\,\left\|\,Q_{\mathcal{D}_n}(x)\right)\right.}, \nonumber
\end{align}
where $\|.\|_{TV}$is the {\it total variation distance} between probability measures, which is given by the $L_1$ norm of the difference between $P(x)$ and $Q_{\mathcal{D}_n}(x)$ as follows:
\begin{align} 
\|P(x)-Q_{\mathcal{D}_n}(x)\|_{TV} &= \|P(x)-Q_{\mathcal{D}_n}(x)\|_{1} \nonumber \\
  &= \int_{\mathcal{X}} |P(x)-Q_{\mathcal{D}_n}(x)|\, dx. 
\label{AAAA2}
\end{align} 
Since the $L_1$ norm is bounded below by the $L_2$ norm, we can lower bound the KL divergence by combining (\ref{AAAA2}) with Pinsker's inequality as follows: 
\begin{align} 
D_{\mathrm {\small KL}}\left(P(x)\,\left\|\,Q_{\mathcal{D}_n}(x)\right)\right. &\geq 2\,\|P(x)-Q_{\mathcal{D}_n}(x)\|^2_{1} \\  
&\geq 2\,\|P(x)-Q_{\mathcal{D}_n}(x)\|^2_{2},
\label{AAAA3}
\end{align} 
and hence it follows that
\begin{align} 
\mathbb{D}_n(\Pi; f_0, f_1,\gamma) &\geq 2\,\mathbb{E}_{x}\left[\,\mathbb{E}_{\mathcal{D}_n}\left[\|P(x)-Q_{\mathcal{D}_n}(x)\|^2_{2}\right]\,\right] \nonumber \\
&= 2\,\mathbb{E}_{x}\left[\,\mathbb{E}_{\mathcal{D}_n}\left[\|T-\mathbb{E}_{\Pi}[\,T\,|\,\mathcal{D}_n\,]\|^2_{2}\right]\,\right], \nonumber 
\end{align}
which leads to the following asymptotic inequality
\begin{align} 
\mathbb{D}_n(\Pi; f_0, f_1,\gamma) \gtrsim \mathbb{E}_{\mathcal{D}_n}\left[\|T-\mathbb{E}_{\Pi}[\,T\,|\,\mathcal{D}_n\,]\|^2_{2}\right].
\label{AAAA4}
\end{align}
By combining (\ref{AAAA4}) and (\ref{AAAA1}), we arrive at (\ref{AAAA0}). From (\ref{AAAA0}), it follows that the optimal information rate is equivalent to the minimax estimation rate in the $L_2(\mathbb{P})$ norm, i.e.
\begin{align} 
I^*_n(\mathcal{F}_0, \mathcal{F}_1,\gamma) \asymp \min_{\hat{T}} \max_{f_0 \in \mathcal{F}_0, f_1 \in \mathcal{F}_1} \mathbb{E}_{\mathcal{D}_n}\left[\|T-\hat{T}\|^2_{2}\right], 
\nonumber
\end{align}
where the estimator $\hat{T}$ is taken to be $\mathbb{E}_{\Pi}[\,T\,|\,\mathcal{D}_n\,]$ since the posterior mean estimator is optimal for the $L_2(\mathbb{P})$ loss. In what follows, we derive the optimal information rate by obtaining the minimax rate of estimation the ITE function $T(x)$.     

Let~$\delta_{\omega}$~be~the~solution~to~$H(\delta_{\omega};\, \mathcal{F}^{\alpha_{\omega}}) \asymp n\,\delta^2_{\omega}$. We will prove that the optimal rate is $\Theta(\delta^2_0 \vee \delta^2_1)$ by first showing that $I^*_n(\mathcal{F}^{\alpha_0},\mathcal{F}^{\alpha_1})$ is lower bounded by, i.e. $I^*_n(\mathcal{F}^{\alpha_0},\mathcal{F}^{\alpha_1}) = \Omega(\delta^2_0 \vee \delta^2_1),$ and then show that $I^*_n(\mathcal{F}^{\alpha_0},\mathcal{F}^{\alpha_1}) = O(\delta^2_0 \vee\delta^2_1)$. We start by observing that the causal inference problem can be described through the following Markov chain 
\begin{align}
(f_0,f_1) \to \mathcal{D}_n \to (\hat{f}_0,\hat{f}_1) \to \hat{T}.\nonumber
\end{align}
The amount of information shared between the true function $T(.)$ and the estimate $\hat{T}(.)$ can be quantified by the {\it mutual information} $I(T;\hat{T})$. Given the Markov chain above, we can upper bound $I(T;\hat{T})$ as follows  
\begin{align}
I(T;\hat{T}) \overset{\small (*)}{\leq} I(T;\mathcal{D}_n) \overset{\small (\star)}{\leq} \sup_{\Pi} I(T;\mathcal{D}_n),
\label{eqA1}
\end{align}
where $(*)$ follows from the {\it data processing} inequality \cite{cover2012elements}, and the supremum in $(\star)$ is taken over all possible priors. $I(T;\hat{T})$ is bounded below by the {\it rate-distortion} function
\begin{align}
I(T;\hat{T}) \geq \inf_{T, \hat{T}:\,\mathbb{E}\|T-\hat{T}\|_2^2 \,\leq\, R_{\Pi}^*} I(T;\hat{T}),
\label{eqA2}
\end{align}
for any $\hat{T}$ satisfying $\mathbb{E}\|T-\hat{T}\|_2^2 \leq R_{\Pi}^*$, where the infimum is taken over all joint distributions of $(T,\hat{T})$. Combining (\ref{eqA1}) and (\ref{eqA2}), we can upper and lower bound the mutual information $I(T;\hat{T})$ as follows
\begin{align} 
\inf_{\,\mathbb{E}\|T-\hat{T}\|_2^2 \,\leq\, R_{\Pi}^*} I(T;\hat{T}) \leq I(T;\hat{T}) \leq \sup_{\Pi} I(T;\mathcal{D}_n).
\label{eqA02}
\end{align}
The lower bound in the chain of inequalities above is intractable, and hence we further lower bound $I(T;\hat{T})$ using {\it Fano's} method \cite{clarke1990information,yang1999information}. That is, we take discrete subsets $\tilde{\mathcal{F}}^{\alpha_0}$ and $\tilde{\mathcal{F}}^{\alpha_1}$ of the function spaces $\mathcal{F}^{\alpha_0}$ and $\mathcal{F}^{\alpha_1}$, and convert the estimation problem to a testing problem. The spaces 
\begin{align}
\tilde{\mathcal{F}}^{\alpha_\omega} = \{\tilde{f}^1_\omega,.\,.\,.,\tilde{f}^{\tilde{M}_\omega}_\omega\},\,\, \tilde{\mathcal{F}}^{\alpha_\omega} \subset \mathcal{F}^{\alpha_\omega},\,\, \omega \in \{0,1\},\nonumber
\end{align}
are constructed such that $\|\tilde{f}^i_\omega-\tilde{f}^j_\omega\| \geq \delta,\, \forall i \neq j$. Let $Q$ be a quantizer that maps elements of $\mathcal{F}^{\alpha_\omega}$ to $\tilde{\mathcal{F}}^{\alpha_\omega},\,\, \omega \in \{0,1\}$. Thus, the causal inference problem can be described through the following Markov chain: 
\begin{align}
(f_0,f_1) \to \mathcal{D}_n \to (\hat{f}_0,\hat{f}_1) \to Q(\hat{f}_0,\hat{f}_1).
\label{eqA3}
\end{align}
Let $\tilde{T} = \tilde{f}^u_1-\tilde{f}^v_0,$ where $\tilde{f}^v_0$ and $\tilde{f}^u_1$ are the functions in $\tilde{\mathcal{F}}^{\alpha_0}$ and $\tilde{\mathcal{F}}^{\alpha_1}$ that are closest to $f_0$ and $f_1$. The discrete element $\tilde{T}$ belongs to a set $\{\tilde{T}^1,.\,.\,.,\tilde{T}^{\tilde{M}_T}\}$, which corresponds to a discretized version of the function space to which $T$ belongs. Using the data processing inequality, we have that  
\begin{align}
I(\tilde{T};\hat{T}) \geq I(\tilde{T};Q(\hat{T})). 
\label{eqA4}
\end{align}
An ``error event" is an event where $Q(\hat{T})$ does not correspond to the true discretized function $\tilde{T}$, i.e. the event $\{\tilde{T} \neq Q(\hat{T})\}$. The error event occurs when
\begin{align}
\|\hat{T}-Q(\hat{T})\| \leq \|\hat{T}-\tilde{T}\|,\, \{\tilde{T} \neq Q(\hat{T})\}.
\label{eqA5}
\end{align}
Thus, the error event implies that $\delta \leq \|Q(\hat{T})-\tilde{T}\|$. Using the triangular inequality, (\ref{eqA5}) can be further bounded as follows: 
\begin{align}
\delta &\leq \|Q(\hat{T})-\tilde{T}\| = \|Q(\hat{T})-\hat{T}+\hat{T}-\tilde{T}\|\nonumber\\ 
&\leq \|Q(\hat{T})-\hat{T}\| + \|\hat{T}-\tilde{T}\|  \nonumber \\
&\leq 2\,\|\hat{T}-\tilde{T}\| \, \implies \|\hat{T}-\tilde{T}\| \geq \frac{\delta}{2}.
\label{eqA6}
\end{align}
Let $P_e$ be the probability of the error event $\{\tilde{T} \neq Q(\hat{T})\}$. From (\ref{eqA6}), $P_e$ can be bounded above as follows 
\begin{align}     
P_e &\defeq \mathbb{P}(\{\tilde{T} \neq Q(\hat{T})\}) \nonumber\\
&= \mathbb{P}(\|Q(\hat{T})-\tilde{T}\| \geq \delta) = \mathbb{P}(\|\hat{T}-\tilde{T}\| \geq \delta/2) \nonumber\\
&= \mathbb{P}(\|\hat{T}-\tilde{T}\|_2^2 \geq \delta^2/4) \nonumber\\
&\overset{\small (\bullet)}{\leq} \frac{4}{\delta^2}\,\mathbb{E}[\|\hat{T}-\tilde{T}\|_2^2] \leq \frac{4}{\delta^2}\,R^*_{\Pi},
\label{eqA7}
\end{align}
where ${\small (\bullet)}$ is an application of Markov's inequality. By combining (\ref{eqA4}) with the result in (\ref{eqA7}), the lower bound in (\ref{eqA02}) can be further bounded below as follows 
\begin{align} 
\inf_{\mathbb{E}\|T-\hat{T}\|_2^2 \,\leq\, R_{\Pi}^*} I(T;\hat{T}) &\geq \inf_{\mathbb{E}\|T-\hat{T}\|_2^2 \,\leq\, R_{\Pi}^*} I(\tilde{T};\hat{T}) \nonumber  \\
&= \inf_{P_e \leq \frac{4}{\delta^2}\,R^*_{\Pi}} I(\tilde{T};\hat{T}) \nonumber \\
&\geq \inf_{P_e \leq \frac{4}{\delta^2}\,R^*_{\Pi}} I(\tilde{T};Q(\hat{T})).\nonumber
\end{align}
The mutual information $I(\tilde{T};Q(\hat{T}))$ can be bounded above as follows
\begin{align} 
I(\tilde{T};Q(\hat{T})) &= I(\tilde{f}_1-\tilde{f}_0;Q(\hat{f}_1-\hat{f}_0))\nonumber \\
&\overset{\small (\odot)}{\leq} I(\tilde{f}_0,\tilde{f}_1;Q(\hat{f}_1-\hat{f}_0))\nonumber \\
&\leq I(\tilde{f}_0,\tilde{f}_1;Q(\hat{f}_0),Q(\hat{f}_1))\nonumber \\
&= I(\tilde{f}_0;Q(\hat{f}_0)) + I(\tilde{f}_1; Q(\hat{f}_1))\nonumber \\
&\leq 2\,\mbox{max}\{I(\tilde{f}_0;Q(\hat{f}_0)), I(\tilde{f}_1;Q(\hat{f}_1))\},
\label{eqA8}
\end{align}
where ${\small (\odot)}$ follows from the data processing inequality. Note that the mutual information $I(\tilde{T};Q(\hat{T}))$ can be written in terms of the KL divergence as \cite{cover2012elements} 
\begin{align} 
I(\tilde{T};Q(\hat{T})) &= D(\mathbb{P}(\tilde{T};Q(\hat{T}))\,||\,\mathbb{P}(\tilde{T}) \cdot \mathbb{P}(Q(\hat{T}))) \nonumber \\
&\geq D(\mbox{Bern}(P_e)\,||\,\mbox{Bern}(1-1/n)) \nonumber \\
&= P_e\log\left(\frac{P_e}{1-1/\tilde{M}_T}\right) + (1-P_e)\log\left(\frac{1-P_e}{1/\tilde{M}_T}\right) \nonumber \\
&= -h(P_e) + \log(\tilde{M}_T)-P_e\,\log(\tilde{M}_T-1) \nonumber \\
&\geq -\log(2) + \log(\tilde{M}_T) - P_e\,\log(\tilde{M}_T),
\label{eqA9}
\end{align}
where $h(.)$ is the binary entropy. From (\ref{eqA9}), we have that 
\begin{align} 
P_e &\geq 1-\frac{I(\tilde{T};Q(\hat{T})) + \log(2)}{\log(\tilde{M}_T)},
\label{eqA10}
\end{align}
which is an incarnation of Fano's inequality. By combining (\ref{eqA8}) with (\ref{eqA10}), we have the following inequality
\begin{align} 
P_e &\geq 1-\frac{I(\tilde{f}_0;Q(\hat{f}_0)) \vee I(\tilde{f}_1;Q(\hat{f}_1)) + \log(\sqrt{2})}{\frac{1}{2}\,\log(\tilde{M}_T)}.
\label{eqA11}
\end{align}
From (\ref{eqA7}), the minimax risk $R^*_{\Pi}$ is bounded below by
\begin{align} 
R^*_{\Pi} &\geq \frac{\delta^2}{4} \left(1-\frac{I(\tilde{f}_0;Q(\hat{f}_0)) \vee I(\tilde{f}_1;Q(\hat{f}_1)) + \log(\sqrt{2})}{\frac{1}{2}\,\log(\tilde{M}_T)}\right).
\nonumber 
\end{align}
The discretization $\tilde{\mathcal{F}}^{\alpha_\omega} = \{\tilde{f}^1_\omega,.\,.\,.,\tilde{f}^{\tilde{M}_\omega}_\omega\}$ corresponds to a $\delta$-packing of the function space $\mathcal{F}^{\alpha_\omega}$, and hence $\tilde{M}_\omega$ is given by the covering number $N(\delta, \mathcal{F}^{\alpha_\omega})$, for $\omega \in \{0,1\}$. It follows that $\tilde{M}_T \geq N(\delta, \mathcal{F}^{\alpha_0}) \vee N(\delta, \mathcal{F}^{\alpha_1})$, and hence we have that
\begin{align} 
R^*_{\Pi} &\geq \frac{\delta^2}{4} \left(1-\frac{I(\tilde{f}_0;Q(\hat{f}_0)) \vee I(\tilde{f}_1;Q(\hat{f}_1)) + \log(\sqrt{2})}{\frac{1}{2}\,\log(N(\delta, \mathcal{F}^{\alpha_0}) \vee N(\delta, \mathcal{F}^{\alpha_1}))}\right).
\nonumber 
\end{align}
The mutual information $I(\tilde{f}_\omega;Q(\hat{f}_\omega))$ can be bounded via the KL divergence as
\begin{align} 
I(\tilde{f}_\omega;Q(\hat{f}_\omega)) &\leq \frac{1}{N^2(\delta, \mathcal{F}^{\alpha_\omega})}\sum_{i,j} D(\mathbb{P}(\tilde{f}^i_\omega)\,||\,\mathbb{P}(\tilde{f}^j_\omega)) \nonumber \\
&\leq 2\,n\,\delta^2.
\nonumber 
\end{align}
Thus, the minimax risk can be bounded below as follows
\begin{align} 
R^*_{\Pi} &\geq \frac{\delta^2}{4} \left(1-\frac{4\,n\,\delta^2 + \log(2)}{\log(N(\delta, \mathcal{F}^{\alpha_0}) \vee N(\delta, \mathcal{F}^{\alpha_1}))}\right),
\nonumber 
\end{align}
and hence we have that 
\begin{align} 
R^*_{\Pi} &\gtrsim \delta^2 - \frac{\delta^4\,n + \delta^2}{\log(N(\delta, \mathcal{F}^{\alpha_0}) \vee N(\delta, \mathcal{F}^{\alpha_1}))}. 
\label{eqA012}
\end{align}
Since $R^*_{\Pi}$ is strictly positive, then we have that 
\begin{align} 
R^*_{\Pi} &\gtrsim \delta^2, \nonumber
\end{align}
where $\delta$ is the solution to the transcendental equation
\begin{align} 
\delta^2 \asymp \frac{\delta^4\,n}{\log(N(\delta, \mathcal{F}^{\alpha_0}) \vee N(\delta, \mathcal{F}^{\alpha_1}))},\nonumber
\end{align}
or equivalently
\begin{align} 
\log(N(\delta, \mathcal{F}^{\alpha_0}) \vee N(\delta, \mathcal{F}^{\alpha_1})) \asymp \delta^2\,n.
\label{eqA12}
\end{align}
The metric entropy of a function space $\mathcal{F}^{\alpha_\omega}$ is given by $H(\delta, \mathcal{F}^{\alpha_\omega}) = \log(N(\delta, \mathcal{F}^{\alpha_\omega})$, and hence (\ref{eqA12}) is written as 
\begin{align} 
H(\delta, \mathcal{F}^{\alpha_0}) \vee H(\delta, \mathcal{F}^{\alpha_1}) \asymp \delta^2\,n.
\label{eqA13}
\end{align}
Since the metric entropy $H(\delta, \mathcal{F}^{\alpha_\omega})$ is a decreasing function of the smoothness parameter $\alpha_\omega$, then it follows that the solution $\delta^*$ of the transcendental equation in (\ref{eqA13}) is given by $\delta^* = \delta_0 \vee \delta_1$, where $\delta_\omega$ is the solution to the equation   
\begin{align} 
H(\delta_{\omega}, \mathcal{F}^{\alpha_\omega}) \asymp \delta_{\omega}^2\,n,\, \omega \in \{0,1\}.
\label{eqA14}
\end{align}
The equation in $(\ref{eqA14})$ has a solution for all $n$ when the function space $\mathcal{F}^{\alpha_\omega}$ has a polynomial or a logarithmic metric entropy \cite{van1998asymptotic}, which is the case for all function spaces of interest (see Table I for evaluations of $\delta_0 \vee \delta_1$ for various function spaces). It follows from (\ref{eqA012}) and (\ref{eqA14}) that
\begin{align} 
R^*_\Pi = \Omega(\delta^2_0 \vee \delta^2_1),\,\, H(\delta_{\omega}, \mathcal{F}^{\alpha_\omega}) \asymp \delta_{\omega}^2\,n, \, \omega \in \{0,1\}, \nonumber
\end{align}
and hence, from (\ref{AAAA0}), we have that
\begin{align} 
I^*_n = \Omega(\delta^2_0 \vee \delta^2_1),\,\, H(\delta_{\omega}, \mathcal{F}^{\alpha_\omega}) \asymp \delta_{\omega}^2\,n, \, \omega \in \{0,1\}. 
\label{eqA15}
\end{align}
We now focus on upper bounding $R^*_\Pi$. From \cite{clarke1990information}, we know that the minimax risk is upper bounded by the channel capacity in (\ref{eqA1}), which is further bounded above by the covering numbers as follows 
\begin{align}
R^*_\Pi \lesssim \frac{1}{n}\left(\,\log(N(\delta,\mathcal{F}^{\alpha_0})) \vee \log(N(\delta,\mathcal{F}^{\alpha_1})) + n\,\delta^2\,\right).\nonumber
\end{align}
For $\delta$ satisfying (\ref{eqA14}), we have that
\begin{align}
\log(N(\delta,\mathcal{F}^{\alpha_0})) \vee \log(N(\delta,\mathcal{F}^{\alpha_1})) = \delta^2\,n,
\nonumber
\end{align}
and hence $R^*_\Pi \lesssim \delta^2_0 \vee \delta^2_1$. It follows that        
\begin{align} 
I^*_n = O(\delta^2_0 \vee \delta^2_1),\,\, H(\delta_{\omega}, \mathcal{F}^{\alpha_\omega}) \asymp \delta_{\omega}^2\,n, \, \omega \in \{0,1\}. 
\label{eqA17}
\end{align}
By combining (\ref{eqA15}) and (\ref{eqA17}), we have that $I^*_n = \Omega(\delta^2_0 \wedge \delta^2_1)$ and $I^*_n = O(\delta^2_0 \vee \delta^2_1)$, and hence it follows that 
\begin{align} 
I^*_n = \Theta(\delta^2_0 \vee \delta^2_1),\,\, H(\delta_{\omega}, \mathcal{F}^{\alpha_\omega}) \asymp \delta_{\omega}^2\,n, \, \omega \in \{0,1\}. 
\label{eqA18}
\end{align}

\section{Proof of Theorem 3}
\renewcommand{\theequation}{\thesection.\arabic{equation}}
Note that when $f_0 \in H^{\alpha_0}_{\mathcal{P}_0}$ and $f_1 \in H^{\alpha_1}_{\mathcal{P}_1}$, the metric entropy of $H^{\alpha_0}_{\mathcal{P}_0}$ and $H^{\alpha_1}_{\mathcal{P}_1}$ are given by \cite{yang1999information}: 
\begin{align}
H(\delta,H^{\alpha_0}_{\mathcal{P}_0}) \asymp \delta^{\frac{-|\mathcal{P}_0|}{\alpha_0}},\,\, H(\delta,H^{\alpha_1}_{\mathcal{P}_1}) \asymp \delta^{\frac{-|\mathcal{P}_1|}{\alpha_1}}.\nonumber
\end{align}
From Theorem 2, we know that the optimal information rate is given by $I^*_n(H^{\alpha_0}_{\mathcal{P}_0},H^{\alpha_1}_{\mathcal{P}_1}) \asymp \delta_0^2 \vee \delta_1^2,$ where $\delta_0$ and $\delta_1$ are the solutions for $\delta^{\frac{-|\mathcal{P}_\omega|}{\alpha_\omega}} \asymp n\,\delta^2_\omega,\, \omega \in \{0,1\}$. Thus, we have that
\begin{align} 
\delta_\omega \asymp n^{\frac{-2\alpha_\omega}{2\alpha_\omega + |\mathcal{P}_\omega|}},\, \omega \in \{0,1\}, \nonumber
\end{align}
and hence the optimal information rate is given by  
\begin{align} 
I^*_n(H^{\alpha_0}_{\mathcal{P}_0},H^{\alpha_1}_{\mathcal{P}_1}) \asymp n^{\frac{-2\alpha_0}{2\alpha_0+|\mathcal{P}_0|}} \vee n^{\frac{-2\alpha_1}{2\alpha_1+|\mathcal{P}_1|}}, \nonumber
\end{align}
From Theorem 1, we know that
\begin{align} 
\mathbb{D}_n(\Pi; f_0, f_1,\gamma) &\asymp \mathbb{E}\left[\,\Vert\,\mathbb{E}_{\Pi}[\,T\,|\,\mathcal{D}_n\,]-T\,\Vert_2^2\,\right]. \nonumber
\end{align}
The $L_2(\mathbb{P})$ loss term on the right hand side can be upper bounded as follows (see Lemma C2): 
\begin{align} 
\mathbb{E}\left[\,\Vert\,\mathbb{E}[\,T\,|\,\mathcal{D}\,]-T\,\Vert_2^2\,\right] &\lesssim \mathbb{E}\left[\,\Vert\,\mathbb{E}[\,f_0\,|\,\mathcal{D}\,]-f_0\,\Vert_2^2\,\right] \nonumber \\
&+ \,\mathbb{E}\left[\,\Vert\,\mathbb{E}[\,f_1\,|\,\mathcal{D}\,]-f_1\,\Vert_2^2\,\right], \nonumber
\end{align}
For a Type-II prior $\Pi \in \Pi_{\beta_{0,1}}^{\circ\circ}$ over the two H\"older spaces $H^{\alpha_0}_{\mathcal{P}_0}$ and $H^{\alpha_1}_{\mathcal{P}_1}$, with $\beta_0 = \alpha_0$ and $\beta_1 = \alpha_1$, the minimax estimation rates for nonparametric regression over $f_0$ and $f_1$ are \cite{stone1982optimal, vaart2011information}
\begin{align} 
\inf_{\hat{f}_0} \sup_{f_0 \in H^{\alpha_0}}\mathbb{E}[\,\|\,\hat{f}_0(\mathcal{D}_n)-f_0\,\Vert_2^2\,] &\asymp n^{\frac{-2\alpha_0}{2\alpha_0+|\mathcal{P}_0|}},\nonumber \\
\inf_{\hat{f}_1} \sup_{f_1 \in H^{\alpha_1}}\mathbb{E}[\,\|\,\hat{f}_1(\mathcal{D}_n)-f_1\,\Vert_2^2\,]  &\asymp n^{\frac{-2\alpha_1}{2\alpha_1+|\mathcal{P}_1|}},\nonumber
\end{align}
and it follows that
\begin{align} 
\mathcal{I}_n(\Pi; f_0, f_1,\gamma) = O\left(n^{\frac{-2\alpha_0}{2\alpha_0+|\mathcal{P}_0|}} \vee n^{\frac{-2\alpha_1}{2\alpha_1+|\mathcal{P}_1|}}\right),\nonumber
\end{align}
which matches the optimal information rate in Theorem 2. Similarly, for a Type-I prior $\Pi \in \Pi_{\beta}^{\circ}$ over a H\"older space $H^{\beta}_{\mathcal{P}_0 \cup \mathcal{P}_1},$ with $\beta = \alpha_0 \wedge \alpha_1$, the term $\mathbb{E}\left[\,\Vert\,\mathbb{E}[\,f_\omega\,|\,\mathcal{D}\,]-f_\omega\,\Vert_2^2\,\right]$ becomes equivalent to the $L_2(\mathbb{P})$ of nonparametric regression of the surface $f : [0,1]^{|\mathcal{P}_0 \cup \mathcal{P}_1|} \times \{0,1\} \to \mathbb{R}$. The minimax estimation rate of such a problem is \cite{stone1982optimal, vaart2011information}
\begin{align} 
\inf_{\hat{f}} \sup_{f \in H^{\alpha_0}}\mathbb{E}[\,\|\,\hat{f}(\mathcal{D}_n)-f\,\Vert_2^2\,] &\asymp n^{\frac{-2(\alpha_0 \wedge \alpha_1)}{2\beta+|\mathcal{P}_0 \cup \mathcal{P}_1|}},
\nonumber
\end{align} 
where the number of feature dimensions $|\mathcal{P}_0 \cup \mathcal{P}_1|$ correspond to all the relevant dimensions for the regression function $f(x,\omega)$. The regression function on the discrete dimension $\omega$ can be estimated at the $\sqrt{n}$ parametric rate and hence it does not affect the minimax estimation rate given above. Since the rate $n^{\frac{-2(\alpha_0 \wedge \alpha_1)}{2\beta+|\mathcal{P}_0 \cup \mathcal{P}_1|}}$ is strictly slower than the optimal rate of $n^{\frac{-2\alpha_0}{2\alpha_0+|\mathcal{P}_0|}} \vee n^{\frac{-2\alpha_1}{2\alpha_1+|\mathcal{P}_1|}}$ for all $\beta > 0$, it follows that 
\begin{align} 
\mathbb{D}_n(\Pi; f_0, f_1,\gamma) \lesssim I^*_n(H^{\alpha_0}_{\mathcal{P}_0},H^{\alpha_1}_{\mathcal{P}_1}),\, \forall \Pi \in \Pi_{\beta}^{\circ},\, \forall \beta > 0.
\nonumber
\end{align} 

\section{Proof of Theorem 4}
\renewcommand{\theequation}{\thesection.\arabic{equation}}
We start by providing three Lemmas which we will use to prove the statement of the Theorem.\\
\\
{\bf Lemma~C1.} Let $\mathcal{X}$ be a compact subset of $\mathbb{R}^d$, $\alpha, \beta \in [0,1]$, and $n, m \in \mathbb{N}_0$. If $n + \beta > m + \alpha$, then $H^{n+\beta}(\mathcal{X})$ is compactly contained in $H^{m+\alpha}(\mathcal{X})$. \,\,\, \IEEEQED \\
\\
{\bf Lemma~C2.} The $L_2(\mathbb{P})$ loss $L = \mathbb{E}\left[\,\Vert\,\mathbb{E}[\,T\,|\,\mathcal{D}_n\,]-T\,\Vert_2^2\,\right]$ is asymptotically bounded above as follows:
\begin{align} 
L &\lesssim \mathbb{E}\left[\,\Vert\,\mathbb{E}[\,f_0\,|\,\mathcal{D}_n\,]-f_0\,\Vert_2^2\,\right] + \mathbb{E}\left[\,\Vert\,\mathbb{E}[\,f_1\,|\,\mathcal{D}_n\,]-f_1\,\Vert_2^2\,\right]. \nonumber
\end{align} 
{\bf Proof.} The $L_2(\mathbb{P})$ loss conditioned on an observational dataset $L(\mathcal{D}_n) = \Vert\,\mathbb{E}[\,T\,|\,\mathcal{D}_n\,]-T\,\Vert_2^2$ is given by:
\begin{align} 
L(\mathcal{D}_n)= \,\Vert\,(\hat{f}_1(x)-\hat{f}_0(x))-(f_1(x)-f_0(x))\,\Vert_2^2, 
\label{awwel1}
\end{align} 
where $\hat{f}_\omega(x) = \mathbb{E}[\,f_\omega(x)\,|\,\mathcal{D}_n\,],$ for $\omega \in \{0,1\}$. The $L_2(\mathbb{P})$ norm in (\ref{awwel1}) can be expressed as follows:
\begin{align} 
L(\mathcal{D}_n) &= \,\Vert\,(\hat{f}_1(x)-\hat{f}_0(x))-(f_1(x)-f_0(x))\,\Vert_2^2, \nonumber \\
&= \int_{\mathcal{X}} ((\hat{f}_1(x)-\hat{f}_0(x))-(f_1(x)-f_0(x)))^2\, d\mathbb{P}(x) \nonumber \\
&= \int_{\mathcal{X}} ((\hat{f}_1(x)-f_1(x))+(f_0(x)-\hat{f}_0(x)))^2\, d\mathbb{P}(x) \nonumber \\
&\leq 2\,\int_{\mathcal{X}} ((\hat{f}_1(x)-f_1(x))^2+(\hat{f}_0(x)-f_0(x))^2)\, d\mathbb{P}(x) \nonumber \\
&= 2\,\int_{\mathcal{X}} (\hat{f}_1(x)-f_1(x))^2\, d\mathbb{P}(x,\omega=1) \nonumber \\
&+ 2\,\int_{\mathcal{X}} (\hat{f}_0(x)-f_0(x))^2\, d\mathbb{P}(x,\omega=0).
\end{align} 
Since $d\mathbb{P}(x,\omega=1) = \gamma(x)\cdot d\mathbb{P}(x)$ and $d\mathbb{P}(x,\omega=0) = (1-\gamma(x))\cdot d\mathbb{P}(x)$, we have that
\begin{align} 
L(\mathcal{D}_n) 
&= 2\,\int_{\mathcal{X}} (\hat{f}_1(x)-f_1(x))^2\, \gamma(x)\cdot d\mathbb{P}(x) \nonumber \\
&+ 2\,\int_{\mathcal{X}} (\hat{f}_0(x)-f_0(x))^2\, (1-\gamma(x))\cdot d\mathbb{P}(x)\nonumber \\
&= 2\,\|\sqrt{\gamma(x)}\cdot(\hat{f}_1(x)-f_1(x))\|^2_{L_2(\mathbb{P})} \nonumber \\
&+ 2\,\|\sqrt{1-\gamma(x)}\cdot(\hat{f}_0(x)-f_0(x))\|^2_{L_2(\mathbb{P})}. \nonumber
\end{align}
Using Cauchy-Schwarz inequality, we obtain the following:
\begin{align}
\|\sqrt{\gamma(x)}(\hat{f}_1(x)-f_1(x))\|^2_2 \leq \|\gamma(x)\|_2 \cdot\|(\hat{f}_1(x)-f_1(x))^2\|_2, \nonumber  
\end{align}
and similarly for $\|\sqrt{1-\gamma(x)}\cdot(\hat{f}_0(x)-f_0(x))\|^2_2$. The proof of the Lemma is concluded by observing that $\|\gamma(x)\|_2$ is $O(1)$ and $\|(\hat{f}_1(x)-f_1(x))^2\|_2 \asymp \|(\hat{f}_1(x)-f_1(x))\|^2_2$. The same result can be arrived at via Minkowski inequality. \,\,\, \IEEEQED \\
\\    
{\bf Lemma~C3.} The support of the prior $\Pi(\beta) = \mathcal{GP}(\mbox{Mat\'ern}(\beta))$ is the space of H\"older functions with order $\beta$.\,\,\, \IEEEQED \\
\\
The~proofs~for~Lemmas~C1~and~C3~are~standard~and~can be found in \cite{evans1997partial} and \cite{vaart2011information} respectively.
  
Recall that the expected Kullback-Leibler risk and the $L_2(\mathbb{P})$ loss are asymptotically equivalent (see Appendix A):
\begin{align}
\mathbb{D}_n(\Pi; f_0, f_1,\gamma) \asymp \mathbb{E}_{\mathcal{D}_n}\left[\,\big\Vert\mathbb{E}_{\Pi}\left[\,T\,|\,\mathcal{D}_n\,\right]-T\big\Vert^2_2\,\right].
\label{eqeqeq1} 
\end{align}
From Lemma C2 and the equivalence in (\ref{eqeqeq1}), we have that
\begin{align}
I_n(\Pi(\beta_0,\beta_1); H^{\alpha_0}, H^{\alpha_1}) &\asymp \mathbb{E}_{\mathcal{D}_n}\left[\,\big\Vert\mathbb{E}_{\Pi}\left[\,T\,|\,\mathcal{D}_n\,\right]-T\big\Vert^2_2\,\right]. \nonumber \\
&\lesssim \mathbb{E}_{\mathcal{D}_n}\left[\,\Vert\,\mathbb{E}_{\Pi}[\,f_0\,|\,\mathcal{D}\,]-f_0\,\Vert_2^2\,\right] \nonumber \\
&+ \mathbb{E}_{\mathcal{D}_n}\left[\,\Vert\,\mathbb{E}_{\Pi}[\,f_1\,|\,\mathcal{D}\,]-f_1\,\Vert_2^2\,\right]. \nonumber
\end{align}
Thus, the information rate achieved by the prior $\Pi(\beta_0,\beta_1)$ is upper bounded by the posterior contraction rates \cite{van2008rates} (rate of convergence of the $L_2(\mathbb{P})$ loss) over the surfaces $f_0$ and $f_1$. For a prior $\mathcal{GP}(\mbox{Mat\'ern}(\beta_\omega))$ and a true function $f_\omega \in H^{\alpha_\omega}$, the contraction rate $\varepsilon^2$ is given by solving the following transcendental equation  \cite{vaart2011information}:    
\begin{align}
\phi_{f_\omega}(\varepsilon) \asymp n\cdot\varepsilon^2,
\label{contraction1}
\end{align}
where $\phi_{f_\omega}(\varepsilon)$ is the {\it concentration function} defined as \cite{van2008rates}:
\begin{align}
\phi_{f_\omega}(\varepsilon) \defeq - \log(\mathbb{P}_{\Pi(\beta_\omega)}(\|f-f_\omega\|_\infty < \varepsilon)). 
\label{contraction2}
\end{align} 
The concentration function measures the amount of prior mass that $\Pi$ places around the true function $f_\omega$. The transcendental equation in (\ref{contraction1}) provides a valid contraction rate whenever consistency holds. Consistency of Bayesian inference holds whenever the true parameter (in this case the true function $f_\omega$) is in the support of the prior \cite{pati2015optimal}. From Lemmas C1 and C3, it follows that the necessary and sufficient conditions for consistency is that $\beta_0 \leq \alpha_0$ and $\beta_1 \leq \alpha_1$.  

In \cite[Lemma 4]{vaart2011information}, the concentration function $\phi_{f_\omega}(\varepsilon)$ for a sufficiently smooth prior $\mathcal{GP}(\mbox{Mat\'ern}(\beta_\omega))$, with $\beta_\omega > d/2$, and a sufficiently smooth true function $f_{\omega} \in H^{\alpha_\omega}$, with $\beta_\omega > d/2$, was obtained as follows: 
\begin{align}
\phi_{f_\omega}(\varepsilon) \lesssim \varepsilon^{-\frac{d}{\beta_\omega}} + \varepsilon^{-\frac{2\beta_\omega-2\alpha_\omega +d}{\alpha_\omega}}.
\label{contraction3}
\end{align}  
Thus, combining (\ref{contraction1}) and (\ref{contraction3}), the posterior contraction rate for $\Pi(\beta_\omega)$ around $f_\omega$ is the solution to:  
\begin{align}
n\cdot\varepsilon^2 \lesssim \varepsilon^{-\frac{d}{\beta_\omega}} + \varepsilon^{-\frac{2\beta_\omega-2\alpha_\omega +d}{\alpha_\omega}},
\label{contraction4}
\end{align}
The solution to (\ref{contraction4}) is given by
\begin{align}
\varepsilon &\lesssim n^{-\frac{\beta_\omega}{2\beta_\omega + d}} + n^{-\frac{\alpha_\omega}{2\beta_\omega+d}}, \nonumber \\
            &\asymp n^{-\frac{(\beta_\omega \wedge \alpha_\omega)}{2\beta_\omega + d}}.
\label{contraction5}
\end{align}
Since consistency holds for $\beta_\omega \leq \alpha_\omega$, then $\varepsilon = O(n^{-\frac{\beta_\omega}{2\beta_\omega + d}})$ and the contraction rate is $n^{-\frac{2\beta_\omega}{2\beta_\omega + d}}$. That is, we can characterize the $L_2(\mathbb{P})$ loss surfaces on $f_0$ and $f_1$ as follows:
\begin{align}
\mathbb{E}_{\mathcal{D}_n}\left[\,\Vert\,\mathbb{E}_{\Pi}[\,f_\omega\,|\,\mathcal{D}\,]-f_\omega\,\Vert_2^2\,\right] \lesssim n^{-\frac{2\beta_\omega}{2\beta_\omega + d}},  
\label{contraction6}
\end{align}
and so it follows that:
\begin{align}
I_n(\Pi(\beta_0,\beta_1); H^{\alpha_0}, H^{\alpha_1}) \lesssim n^{-\frac{2\beta_0}{2\beta_0 + d}} + n^{-\frac{2\beta_1}{2\beta_1 + d}} , \nonumber
\end{align}
which concludes the proof of the Theorem. 

\section{Proof of Theorem 5}
\renewcommand{\theequation}{\thesection.\arabic{equation}}
The empirical smoothness estimate $\hat{\beta}_n$ is obtained by minimizing the empirical objective:
\begin{align}
L(\mathcal{D}_n, \beta) = \mathbb{E}_{f_0,f_1 \sim d\Pi_\beta(.\,|\,\mathcal{D}_n)}[\,\mathbb{D}_n(\Pi_\beta; f_0, f_1)\,|\,\mathcal{D}_n\,]. 
\label{F1}
\end{align}
We optimize (\ref{F1}) via model selection with $J$-fold cross-validation. From Theorem 4, we know that for a large sample ($n \uparrow \infty$), the true loss function $L(\beta) = \mathbb{E}_{\mathcal{D}_n}[L(\mathcal{D}_n, \beta)]$ has a unique minimizer: $\beta^* = (\alpha_0 \wedge \alpha_1)$. Let the set $\{\beta^{(1)},.\,.\,.,\beta^{(K_n)}\}$ be a set of $K_n$ candidate minimizers (smoothness levels) of the true loss $L$. Let $B_n = \{B_n(i)\}_i \in \{0,1\}^n$ be a binary split vector which allocates every data point $i$ in $\mathcal{D}_n$ to either of the training or validation sets. We define $\mathbb{P}^T_{n,B_n}$ and $\mathbb{P}^V_{n,B_n}$ be the empirical distributions of the training and validation sets, and let $v$ be the fraction of data allocated to the validation set. The empirical cross-validated risk estimate is defined as
\begin{align}
L^{k_n}(\mathcal{D}_n, \beta^{(k_n)}) &= \mathbb{E}_{B_n} \int L(\mathcal{D}_n,\mathbb{P}^T_{n,B_n}, \beta^{(k_n)})\, d\mathbb{P}^V_{n,B_n} \nonumber \\
&= \mathbb{E}_{B_n} \mbox{\footnotesize $\frac{1}{\sum_{B_n(i)}}$} \mbox{\footnotesize $\sum_{\{i : B_n(i) = 1\}}$}\, L(\mathcal{D}^i_n,\mathbb{P}^T_{n,B_n}). \nonumber
\end{align}
The candidate in $\{\beta^{(1)},.\,.\,.,\beta^{(K_n)}\}$ that minimizes the cross-validated risk $L^{k_n}(\mathcal{D}_n)$ is 
\begin{align}
\hat{k}^*_{n} \defeq \arg \min_{k_n} L(\mathcal{D}_n,\beta^{(k_n)}).  
\label{F3}
\end{align}
The consistency of the estimator $\beta^{(K^*_n)}$ follows from the results of Dudoit and van der Laan on the asymptotic performance of model selection via cross-validation for general loss functions \cite{dudoit2005asymptotics}. Suppose that $\sup_{\mathcal{D}_n,\beta} L(\mathcal{D}_n,\beta) \leq \infty$, $\beta^* \in \{\beta^{(1)},.\,.\,.,\beta^{(K_n)}\}$, and $\log(K_n)/(\sqrt{nv}(L^{k_n}-L)) \overset{p}{\to} 0$ as $n \to \infty$. Then, from Theorem 2 in \cite{dudoit2005asymptotics}, we have that $L^{k_n}(\mathcal{D}_n, \beta^{(k_n)})-L(\mathcal{D}_n,\beta) \overset{p}{\to} 0$. Since $\beta^* = (\alpha_0 \wedge \alpha_1)$ is a unique minimizer of $L(\mathcal{D}_n,\beta)$, then it follows from the {\it argmin continuous mapping} theorem for $M$-estimators that $\beta^{(K^*_n)} \to (\alpha_0 \wedge \alpha_1)$ \cite{van1998asymptotic}.   

\section{Proof of Theorem 6}
\renewcommand{\theequation}{\thesection.\arabic{equation}}
Using the asymptotic equivalence between the KL and the $L_2(\mathbb{P})$ risks, we have that 
\begin{align}
\hat{\beta}_n &= \arg \min_{\beta} \mathbb{E}_{\Pi(.\,|\,\mathcal{D}_n,\beta)}[\,\mathbb{E}_{x}[\,D_{\mathrm {\small KL}}(P(x)\,\|\,Q_{\mathcal{D}_n}(x))\,]\,] \nonumber \\
&\asymp \arg \min_{\beta} \mathbb{E}_{\Pi(.\,|\,\mathcal{D}_n,\beta)}[\,\mathbb{E}_{x}[\,\|\mathbb{E}_{\Pi}[T\,|\,\mathcal{D}_n]-T\|^2_2\,]\,]. \nonumber 
\end{align} 
Since there is a unique minimizer for both the KL and $L_2(\mathbb{P})$ risks when the number of samples is asymptotically large, we work on the $L_2(\mathbb{P})$ risk objective to obtain the optimal solution. The posterior Bayesian risk $R(\theta,{\bf \hat{f}}; \mathcal{D})$ for a point estimate ${\bf \hat{f}}$ is given by        
\begin{align}
R(\theta,{\bf \hat{f}}; \mathcal{D}) = \mathbb{E}_{\theta}\left[\left.\hat{\mathcal{L}}({\bf \hat{f}};{\bf K}_\theta,{\bf Y^{(W)}},{\bf Y^{(1-W)}})\,\right|\,\mathcal{D}\right], \nonumber
\end{align}
where the expectation in is taken with respect to ${\bf Y^{(1-W)}}|\mathcal{D}$. The Bayesian risk can be written as
\begin{align}
R(\theta,{\bf \hat{f}}; \mathcal{D}) = \int \hat{\mathcal{L}}({\bf \hat{f}};{\bf K}_\theta,{\bf Y^{(W)}},{\bf Y^{(1-W)}})\,d\mathbb{P}_{\theta}({\bf Y^{(1-W)}}|\mathcal{D}). \nonumber
\end{align}
The loss function $\hat{\mathcal{L}}$ conditional on a realization of the counterfactual outcomes is given by 
\begin{align}
&\hat{\mathcal{L}}({\bf \hat{f}};{\bf K}_{\theta},{\bf Y^{(W)}},{\bf Y^{(1-W)}}) = \nonumber \\
&\frac{1}{n}\sum_{i=1}^{n} \left(\hat{\bf f}^T(X_i){\bf e}-(1-2W_i)\left(Y^{(1-W_i)}_i-Y^{(W_i)}_i\right)\right)^{2}. \nonumber
\end{align}

\begin{figure*}[!t]
\normalsize
\setcounter{mytempeqncnt}{\value{equation}}
\setcounter{equation}{31}
\begin{align}
\label{A7}
(\hat{f}^*,\theta^*) &= \arg \min_{\hat{f}, \theta} \int \frac{1}{n}\sum_{i=1}^{n} \left(\hat{\bf f}^T(X_i){\bf e}-(1-2W_i)\left(Y^{(1-W_i)}_i-Y^{(W_i)}_i\right)\right)^{2}d\mathbb{P}_{\theta}(Y_i^{(1-W)}|\mathcal{D}).\\
\theta^* &= \arg \min_{\theta} \int \frac{1}{n}\sum_{i=1}^{n} \left(\mathbb{E}_{\theta}[{\bf f}^T(X_i)\,|\,\mathcal{D}]{\bf e}-(1-2W_i)\left(Y^{(1-W_i)}_i-Y^{(W_i)}_i\right)\right)^{2}d\mathbb{P}_{\theta}(Y_i^{(1-W)}|\mathcal{D}).\\
R &= \frac{1}{n}\sum_{i=1}^{n} \int \left((1-2W_i)\left((Y^{(W_i)}_i-\mathbb{E}_{\theta}[f_{W_i}(X_i)\,|\,\mathcal{D}])-(Y^{(1-W_i)}_i-\mathbb{E}_{\theta}[f_{1-W_i}\,|\,\mathcal{D}])\right)\right)^{2}\, d\mathbb{P}_{\theta}(Y_i^{(1-W)}|\mathcal{D}).\\
R &= \frac{1}{n}\sum_{i=1}^{n} \underbrace{\int (Y^{(W_i)}_i-\mathbb{E}_{\theta}[f_{W_i}(X_i)\,|\,\mathcal{D}])^2 \, d\mathbb{P}_{\theta}(Y_i^{(1-W)}|\mathcal{D})}_{R_1} + \underbrace{\int (Y^{(1-W_i)}_i-\mathbb{E}_{\theta}[f_{1-W_i}\,|\,\mathcal{D}])^{2} \, d\mathbb{P}_{\theta}(Y_i^{(1-W)}|\mathcal{D})}_{R_2} \nonumber \\
& - \underbrace{2 \int (Y^{(W_i)}_i-\mathbb{E}_{\theta}[f_{W_i}(X_i)\,|\,\mathcal{D}]) \, (Y^{(1-W_i)}_i-\mathbb{E}_{\theta}[f_{1-W_i}\,|\,\mathcal{D}]) \, d\mathbb{P}_{\theta}(Y_i^{(1-W)}|\mathcal{D})}_{R_3}.  
\end{align}
\setcounter{equation}{\value{mytempeqncnt}+4}
\hrulefill
\vspace*{4pt}
\end{figure*}
The optimal hyper-parameter and interpolant $(\hat{f}^*,\theta^*)$ are obtained through the following optimization problem in (F.1). The optimization problem can solved separately for $\theta$ and $\hat{f}$; we know from Theorem 1 that for any given $\theta$, the optimal interpolant $\hat{{\bf f}} = \mathbb{E}_{\theta}[{\bf f}\,|\,\mathcal{D}]$. Hence, the optimal hyper-parameter $\theta^*$ can be found by solving the optimization problem in (F.2). The objective function $R$ can thus be written as in (F.3) and further reduced as in (F.4).
 
Note that since $Y_i^{(W_i)} = f_{W_i}(X_i) + \epsilon_{i,W_i}$, then we have that $\mathbb{E}_{\theta}[f_{W_i}(X_i)\,|\,\mathcal{D}] = \mathbb{E}_{\theta}[Y_i^{(W_i)}\,|\,\mathcal{D}]$ and $\mathbb{E}_{\theta}[f_{1-W_i}(X_i)\,|\,\mathcal{D}] = \mathbb{E}_{\theta}[Y_i^{(1-W_i)}\,|\,\mathcal{D}]$. Therefore, we can evaluate the terms $R_1$, $R_2$ and $R_3$ as follows
\begin{align}
R_1 &= \frac{1}{n}\sum_{i=1}^{n} \int (Y^{(W_i)}_i-\mathbb{E}_{\theta}[f_{W_i}(X_i)\,|\,\mathcal{D}])^2 \, d\mathbb{P}_{\theta}(Y_i^{(1-W)}|\mathcal{D}) \nonumber \\
&= \frac{1}{n}\sum_{i=1}^{n} \int (Y^{(W_i)}_i-\mathbb{E}_{\theta}[Y^{(W_i)}_i\,|\,\mathcal{D}])^2 \, d\mathbb{P}_{\theta}(Y_i^{(1-W)}|\mathcal{D}) \nonumber \\
&= \frac{1}{n} \|{\bf Y^{(W)}}-\mathbb{E}_{\theta}[{\bf f}\,|\,\mathcal{D}]\|_2^2,
\end{align} 
and
\begin{align}
R_2 &= \frac{1}{n}\sum_{i=1}^{n} \int (Y^{(1-W_i)}_i-\mathbb{E}_{\theta}[f_{1-W_i}\,|\,\mathcal{D}])^{2} \, d\mathbb{P}_{\theta}(Y_i^{(1-W)}|\mathcal{D}) \nonumber \\
&= \frac{1}{n}\sum_{i=1}^{n} \int (Y^{(1-W_i)}_i-\mathbb{E}_{\theta}[Y^{(1-W_i)}_i\,|\,\mathcal{D}])^{2} \, d\mathbb{P}_{\theta}(Y_i^{(1-W)}|\mathcal{D}) \nonumber \\
&= \frac{1}{n}\sum_{i=1}^{n} \mbox{Var}[\, Y^{(1-W_i)}_i\,|\,\mathcal{D}\,], \nonumber \\
&= \frac{1}{n} \|\mbox{Var}[\, {\bf Y^{(1-W)}}\,|\,\mathcal{D}\,]\|_1,
\end{align} 
and
\begin{align}
R_3 = &\frac{1}{n}\sum_{i=1}^{n} \int (Y^{(W_i)}_i-\mathbb{E}_{\theta}[f_{W_i}\,|\,\mathcal{D}]) \nonumber \\
      &(Y^{(1-W_i)}_i-\mathbb{E}_{\theta}[f_{1-W_i}\,|\,\mathcal{D}]) \, d\mathbb{P}_{\theta}(Y_i^{(1-W)}|\mathcal{D}) = 0 \nonumber
\end{align} 
Therefore, $\theta^*$ is found by minimizing $\|{\bf Y^{(W)}}-\mathbb{E}_{\theta}[{\bf f}\,|\,\mathcal{D}]\|_2^2+\|\mbox{Var}[\, {\bf Y^{(1-W)}}\,|\,\mathcal{D}\,]\|_1$.

\bibliographystyle{IEEEtran}
\bibliography{IEEErefs}

\end{document}